\documentclass[seceq]{ptptex}

\renewcommand{\Re}{\text{Re}}
\renewcommand{\Im}{\text{Im}}
\pubinfo{Vol.~116, No.~1, July 2006}
\notypesetlogo
\title{%
Theory of Non-Equilibirum States \\
Driven by  Constant Electromagnetic Fields}

\subtitle{Non-Commutative Quantum Mechanics in the Keldysh Formalism}

\author{%  
Shigeki \textsc{Onoda},$^{1,}$\footnote{E-mail: sonoda@appi.t.u-tokyo.ac.jp}
Naoyuki \textsc{Sugimoto}$^2$ and Naoto \textsc{Nagaosa}$^{3,4}$}

\inst{%    
$^1$Spin Superstructure Project, ERATO, Japan Science and Technology Agency,\\ c/o Department of Applied Physics, University of Tokyo, Tokyo 113-8656, Japan\\
$^2$Department of Applied Physics, University of Tokyo, Tokyo 113-8656, Japan\\
$^3$CREST, Department of Applied Physics, University of Tokyo, Tokyo 113-8656\\
$^4$Correlated Electron Research Center, National Institute of Advanced Industrial Science and Technology, Tsukuba 305-8562, Japan}

\recdate{May 14, 2006}%   

\abst{%      
We develop a general theory of non-equilibrium states based on 
the Keldysh formalism, in particular, for charged-particle systems 
under static uniform electromagnetic fields. The Dyson equation for 
the uniform stationary state is rewritten in a compact gauge-invariant 
form by using the Moyal product in the phase space of energy-momentum 
variables, whcich originally do not commute in the case of the 
conventional operator algebra. Expanding the Dyson equation in 
electromagnetic fields, a systematic method for the order-by-order
calculation of linear and non-linear responses from the zeroth-order 
Green's functions is obtained. In particular, we find that for impurity 
problems, up to linear order in the electric field, the present approach
provides a diagrammatic method for the St\u{r}eda formula. This approach 
also generalizes the semi-classical Boltzmann transport theory to fully 
quantum-mechanical and/or multi-component systems. In multi-component 
systems and/or for Hall transport phenomena, however, this quantum 
Boltzmann transport theory, constructed from the anti-symmetric 
combination of two different representations for the Dyson equation, 
does not uniquely specify the non-equilibrium state, but the symmetric 
combination is required. We demonstate the formalism to calculate 
longitudinal and Hall electric conductivities in an isotropic 
single-band electron system in the clean limit. It is found that 
the results are fully consistent with those obtained by Mott and 
Ziman in terms of the semi-classical Boltzmann transport theory.
}

\begin{document}
\maketitle

\section{Introduction}
\label{sec:intro}

  Theory of non-equilibrium states is one of the most fundamental and important topics in statistical and condensed-matter physics. For instance, applying an electric field drives a charged system off the equilibrium. Electron transport properties in a condensed matter under an applied electric/magnetic field are conventionally described in terms of the semi-classical Boltzmann transport theory~\cite{Mott67,Ziman67}. Rigorous formulations based on quantum mechanics have also been developed by several authors. The theory of the linear response of equilibrium systems to external perturbations was established by Kubo~\cite{Kubo57} and have also been formulated in various manners~\cite{Nakajima58,Zwanzig61,Mori65}. In this case, the linear responses are described by correlation functions of the equilibrium state. To go beyond the linear response in the Kubo formalism, one needs to expand the correlation function further in the external perturbation. From this viewpoint, Fukuyama and coworkers performed a diagrammatic calculation of the Hall effect in metals in the presence of impurities~\cite{Fukuyama69_1,Fukuyama69_2}, which has also been generalized for Fermi liquids~\cite{KohnoYamada88}.

  The Keldysh~\cite{Keldysh}, or Baym-Kadanoff~\cite{KadanoffBaym}, formalism is a powerful theoretical framework to treat non-linear and/or non-equilibirum phenomena~\cite{Altshuler78,RammerSmith86,Mahan} by directly handling the two-point Green's functions of all the independent analyticities. In particular, in the Wigner representation in terms of the center-of-mass time-space coordinate $X^\mu$ and the relative canonical energy-momentum $p^\mu$, each product appearing in the Dyson equation for a uniform equilibrium, i.e., $X^\mu$-independent, system is replaced by the so-called Moyal product~\cite{Moyal49} in this ``phase'' space, in order to fully describe the convolutions of $X^\mu$-dependent Green's functions~\cite{RammerSmith86}.

  It is known that taking an antisymmetric combination of two different representations for the Dyson equation and then performing the gradient expansion, the Keldysh formalism gives a fully quantum-mechanical generalization of the semi-classical Boltzmann transport theory~\cite{Altshuler78,RammerSmith86,Mahan}, which is often referred to as the quantum Boltzmann transport theory. In particular, for uniform stationary states under constant, i.e., uniform and static, electromagnetic fields, the Green's functions can be expressed as apparently gauge-invariant functionals of the energy-momentum variable $\pi^\mu=p^\mu-qA^\mu(X)$. Here, $q$ is the electric charge of the particle and $A^\mu(X)$ is the scalar and vector potential of the electromagnetic field. This $\pi^\mu$ is the Wigner representation of the mechanical or kinetic momentum~\cite{Altshuler78} or the covariant derivative, which should not be confused with the canonical momentum $p^\mu$.

  For open systems, the Keldysh formalism is combined with the Landauer-Buttiker formalism~\cite{Buttiker94,LandauerButtiker} by imposing a certain boundary condition, or one can deal with the time-dependent Schr\"{o}dinger equation to observe the time evolution of a non-equilibrium state~\cite{Frensley90}.
  In closed systems, however, it is necessary to develop systematic ways of solving non-linear and/or non-equilibrium problems without any semi-claasical approximation. Not even the linear response in multi-component systems and Hall effects can be described solely by the quantum Boltzmann equation, even though it is correctly generalized for these cases. Instead, in addition to the quantum Boltzmann transport equation, one must directly handle the Dyson equation under the external fields in multi-component systems~\cite{HaugJauho}. This point was recently demonstrated in the context of the spin Hall effect~\cite{LiuLei,Sugimoto} and the anomalous Hall effect~\cite{Onoda_ahe}. It is remarkable that the semi-classical Boltzmann equation is not sufficient for explaining the Hall transport phenomena.

  In this paper, we develop a generic theoretical framework for systematically calculating the linear and the non-linear responses to an applied uniform and steady electromagnetic field. We obtain a compact form for the Dyson equation in terms of the Moyal product in the $\pi$ space in the presence of the electromagnetic field tensor $F_{\mu\nu}$, which re-expresses the non-commutativity of the components of the mechanical momentum in the conventional operator algebra. This yields a master equation for solving non-linear and/or non-equilibrium but uniform stationary problems. Up to the linear response to the electric field, we obtain an expression for the electric conductivity in the form of the St\u{r}eda formula~\cite{SmrckaStreda77,Streda82}, which was originally intended for non-interacting electrons. Now we reveal the diagrammatic technique in the St\u{r}eda formula for more generic cases. Generalization to non-uniform and/or time-dependent problems is also straightforward.

  We demonstrate the formalism by applying it to the conductivity tensor, including the longitudinal conductivity and the transverse Hall conductivity in an isotropic single-band electron model in the presence of impurities. In the clean limit, our results coincide with those of the semi-classical Boltzmann transport theory~\cite{Mott67,Ziman67}. We also found the correspondence to diagrammatic calculations of vertex corrections due to the impurity potential in terms of the Kubo formula in the case of the linear-response longitudinal conductivity~\cite{AGD}. 

Major advantages of the present formalism are the following.
\begin{enumerate}
\item The Green's functions can be described in terms of gauge-invariant functionals of the energy-momentum $\pi^\mu=p^\mu-qA^\mu(X)$ in an arbitrary order in the electromagnetic fields. Therefore, the gauge invariance of physical quantities after integration over $\pi$ is evident from the beginning.
\item It allows for the electric current to be expressed as a sum of the equilibrium thermodynamic part, which is proportional to the Bose or Fermi distribution function $f_\mp(\varepsilon)$, and the nonequilibrium parts given by the energy ($\varepsilon$) derivatives $\partial_\varepsilon^m f_\mp(\varepsilon)$ ($m=1,\cdots,n$) in the case of the $n$-th-order response. This enables us to carry out a diagrammatic calculation of the St\u{r}eda formula~\cite{Streda82} and its generalization to non-linear responses.
\item It is easy to treat multi-band/component systems. In general, not only the magnetic field but also the electric field alters the retarded and advanced Green's functions, even in the absence of an interaction. This is beyond the scope of the conventional quantum Boltzmann theory, but it can be systematically calculated in the present approach.
\end{enumerate}

This paper is organized as follows. In \S\ref{sec:Keldysh}, we briefly
introduce the Keldysh formalism for non-equilibrium Green's functions. Then,
we formulate a generic framework to systematically calculate the bulk
quantum distribution function in response to the uniform static
electromagnetic fields for generic multi-component bosonic and fermionic
systems. Sections~\ref{sec:1st}--\ref{sec:transport_1} are devoted to
explicit applications of this generic formalism. In \S\ref{sec:1st}, the
linear response theory is derived. In \S\ref{sec:2nd}, it is generalized to
the cross terms of the electric and magnetic fields. Then, in
\S\ref{sec:transport_1}, we study a simple model of an isotropic single-band
electron system as a concrete application of the present formalism. The
equivalence to the Kubo formula and the St\u{r}eda formula and the merits of
this approach in practical calculations are demonstrated. We have also
obtained results for the longitudinal and the Hall conductivities that are
consistent with the semi-classical ones obtained by Mott~\cite{Mott67} and
Ziman~\cite{Ziman67}. Section \ref{sec:discuss} is devoted to discussion and conclusions.

\section{Keldysh formalism for non-equilibrium Green's functions}
\label{sec:Keldysh}

To consider responses of a charged-particle system to external electromagnetic fields, we employ the Keldysh formalism~\cite{RammerSmith86,Mahan} for non-equilibrium Green's functions. Here, in addition to the retarded and advanced Green's functions, one introduces the Keldysh or the lesser Green's function. In particular, the lesser Green's function plays the role of the quantum distribution function. These non-equilibrium and/or non-linear Green's functions can be calculated at least by performing an order-by-order expansion in the electromagnetic fields.

\subsection{Model}
\label{subsec:model}

We consider a generic multi-component model of bosons or fermions with an electric charge $q$ whose Hamiltonian is given by
\begin{equation}
  \hat{H}_0^{\rm tot}(\mib{x},\mib{p})=\hat{H}_0(\mib{p})+\hat{V}_0(\mib{x}),
  \label{eq:H_0^tot}
\end{equation}
with the space coordinate $\mib{x}=(x,y,z)$ and the conjugate momentum $\mib{p}=-i\hbar\mib{\nabla}_x$. Here, $\hat{H}_0(\mib{p})$ and $\hat{V}_0(\mib{x})$ denote the non-interacting and the impurity-potential parts of the Hamiltonian, respectively. Henceforth, a variable with a hat represents a matrix in the space of internal degrees of freedom including spins.

Next, we introduce the external electromagnetic field $F^{\mu\nu}=\partial_{x_\mu} A^\nu(x)-\partial_{x_\nu} A^\mu(x)$. Henceforth, $x^\mu=(t,\mib{x})$ and $x_\mu=(-t,\mib{x})$ denote the time and space coordinates, $\partial_{x^\mu}=(\partial_t,\mib{\nabla}_x)$ and $\partial_{x_\mu}=(-\partial_t,\mib{\nabla}_x)$ the derivatives, and $A^\mu(x)=(\phi(x),\mib{A}(x))$ and $A_\mu(x)=(-\phi(x),\mib{A}(x))$ the sets of scalar and vector potentials yielding the electromagnetic field tensor $F_{\mu\nu}(x)=\partial_{x^\mu}A_\nu(x)-\partial_{x^\nu}A_\mu(x)$. Then, Eq.~(\ref{eq:H_0^tot}) is modified by $A(x)$ as
\begin{eqnarray}
  \hat{H}^{\rm tot}(x,\mib{p}) &=& \hat{H}^{\rm tot}_0(x,\mib{p}-q\mib{A}(x))+q\phi(x)
  \nonumber\\
  &=&\hat{H}_0(\mib{p}-q\mib{A}(x))+\hat{V}_0(\mib{x})+q\phi(x).
  \label{eq:H^tot}
\end{eqnarray}
Although we have included neither the Coulomb interaction among the electrons nor the phonons into the Hamiltonian, the following formalism based on the Dyson equation holds even in the presence of the Coulomb interaction and the phonons if the corresponding self-energy corrections are properly included according to the Feymann rule~\cite{RammerSmith86}.

\subsection{Green's functions}
\label{subsec:Green}

The retarded, advanced, Keldysh and lesser Green's functions are defined as
\begin{subequations}
\begin{eqnarray}
  \left\{\hat{G}^R\right\}_{(\alpha_1;\alpha_2)}(x_1;x_2)
  &\equiv&-i\theta(t_1-t_2)\Bigl\langle\left[\Psi_{\alpha_1}(x_1),\Psi_{\alpha_2}^\dagger(x_2)\right]_{\mp}\Bigr\rangle,
  \label{eq:G^R:def}\\
  \left\{\hat{G}^A\right\}_{(\alpha_1;\alpha_2)}(x_1;x_2)
  &\equiv&\ \ i\theta(t_2-t_1)\Bigl\langle\left[\Psi_{\alpha_1}(x_1),\Psi^\dagger_{\alpha_2}(x_2)\right]_\mp\Bigr\rangle,
  \label{eq:G^A:def}\\
  \left\{\hat{G}^K\right\}_{(\alpha_1;\alpha_2)}(x_1;x_2)
  &\equiv&-i\Bigl\langle\left[\Psi_{\alpha_1}(x_1),\Psi^\dagger_{\alpha_2}(x_2)\right]_\pm\Bigr\rangle,
  \label{eq:G^K:def}\\
  \left\{\hat{G}^<\right\}_{(\alpha_1;\alpha_2)}(x_1;x_2) &\equiv& \mp i\Bigl\langle\Psi^\dagger_{\alpha_2}(x_2)\Psi_{\alpha_1}(x_1)\Bigr\rangle
  \nonumber\\
  &=& \frac{1}{2}\left(-\hat{G}^R(x_1;x_2)+\hat{G}^A(x_1;x_2)+\hat{G}^K(x_1;x_2)\right),
  \label{eq:G^<}
\end{eqnarray}%
\end{subequations}
respectively. Here, the upper and lower signs correspond to bosons and fermions, respectively. $\Psi_{\alpha_m}^\dagger(x_m)$ and $\Psi_{\alpha_m}(x_m)$ represent the creation and annihilation operators of a boson or a fermion with the internal degree of freedom labeled by $\alpha_m$ at the time-space coordinate $x_m$.

In the present approach, the lesser Green's function represents the electron distribution function (multiplied by $2\pi i$) even in the presence of the external fields and is defined for the full time-space variables. This is in sharp contrast to the situation in the semi-classical Boltzmann transport theory, where only its simultaneous component with $t_1=t_2$ is considered. In the semi-classical case, quasiparticles are implicitly assumed to be well-defined so that the broadening of the peak in their spectra is negligibly small compared with the other energy scales. By contrast, the present approach is applicable even in the case of a strong scattering amplitude for which the usual elastic approximation and thus the semi-classical Boltzmann transport theory become invalid. 

\subsection{Dyson equations}
\label{subsec:Dyson}

In the conventional Larkin-Ovchinnikov representation, one introduces the following matrix-form Green's function and self-energy:
\begin{eqnarray}
  \underline{\hat{G}}(x_1;x_2) &=&
  \left(\begin{array}{cc}
  \hat{G}^R(x_1;x_2) & \hat{G}^K(x_1;x_2)\\
  0                  & \hat{G}^A(x_1;x_2)
  \end{array}\right),
  \label{eq:_G_:LO}\\
  \underline{\hat{\Sigma}}(x_1;x_2) &=&
  \left(\begin{array}{cc}
  \hat{\Sigma}^R(x_1;x_2) & \hat{\Sigma}^K(x_1;x_2)\\
  0                       & \hat{\Sigma}^A(x_1;x_2)
  \end{array}\right).
  \label{eq:_Sigma_:LO}
\end{eqnarray}
Here, $\delta(x_1-x_2)$ represents $\delta(t_1-t_2)\delta(\mib{x}_1-\mib{x}_2)$, and $\hat{\Sigma}^{R(A)}(x_1;x_2)$ and $\hat{\Sigma}^{K(<)}(x_1;x_2)$ denote the retarded (advanced) and Keldysh (lesser) self-energies, respectively. In the present paper, however, we further perform an orthonormal transformation in terms of 
\begin{equation}
  \underline{O}=\left(\begin{array}{cc}
  1 & 1\\
  0 & 1
  \end{array}\right)
\label{eq:O}
\end{equation}
as
\begin{eqnarray}
  \underline{\hat{G}}(x_1;x_2)&\leftarrow&\underline{O}\ \underline{\hat{G}}(x_1;x_2)\underline{O}^{-1}=\left(\begin{array}{cc}
  \hat{G}^R(x_1;x_2) &2\hat{G}^<(x_1;x_2)\\
  0                  & \hat{G}^A(x_1;x_2)
  \end{array}\right),
  \label{eq:_G_}\\
  \underline{\hat{\Sigma}}(x_1;x_2)&\leftarrow&\underline{O}\ \underline{\hat{\Sigma}}(x_1;x_2)\underline{O}^{-1}=
  \left(\begin{array}{cc}
  \hat{\Sigma}^R(x_1;x_2) &2\hat{\Sigma}^<(x_1;x_2)\\
  0                       & \hat{\Sigma}^A(x_1;x_2)
  \end{array}\right).
  \label{eq:_Sigma_}
\end{eqnarray}
This representation has the advantage of allowing the direct calculation of the lesser component without any manipulation of the Keldysh components.

According to Eq.~(\ref{eq:H^tot}), we can write the equations of motion or the Dyson equations for the Green's function in terms of the gauge covariant derivative $\overrightarrow{D}_{{x_m}_\mu}=\overrightarrow{\partial}_{{x_m}_\mu}-iqA^\mu(x_m)/\hbar$ and its conjugate $\overleftarrow{D}^*_{{x_m}_\mu}=\overleftarrow{\partial}_{{x_m}_\mu}+iqA^\mu(x_m)/\hbar$ as
\begin{subequations}
\begin{eqnarray}
  i\hbar \overrightarrow{D}_{t_1}\underline{\hat{G}}(x_1;x_2)-\hat{H}_0(-i\hbar \overrightarrow{\mib{D}}_{x_1})\underline{\hat{G}}(x_1;x_2)-(\underline{\hat{\Sigma}}*\underline{\hat{G}})(x_1;x_2)&=&\delta(x_1-x_2),
\ \ \ \ \ 
  \label{eq:Dyson1_0}\\
  -i\underline{\hat{G}}(x_1;x_2)\hbar \overleftarrow{D}^*_{t_2}-\underline{\hat{G}}(x_1;x_2)\hat{H}_0(i\hbar \overleftarrow{\mib{D}}^*_{x_2})-(\underline{\hat{G}}*\underline{\hat{\Sigma}})(x_1;x_2)&=&\delta(x_1-x_2).
\ \ \ \ \ 
  \label{eq:Dyson2_0}
\end{eqnarray}
\end{subequations}
Henceforth, $\overrightarrow{\partial}$ and $\overleftarrow{\partial}$ denote the derivatives operating on the right-hand and left-hand sides, respectively, and $*$ represents the convolution, defined by
\begin{equation}
  (\underline{\hat{\Upsilon}}*\underline{\hat{\Xi}})(x_1;x_2) = \int\!dx_3\underline{\hat{\Upsilon}}(x_1;x_3)\underline{\hat{\Xi}}(x_3;x_2),
  \label{eq:convolution:x_1;x_2}
\end{equation}
where $\int\!dx$ represents $\int\!dt\int\!d\mib{x}$. 
The Dyson equations (\ref{eq:Dyson1_0}) and (\ref{eq:Dyson2_0}) can also be written in the forms
\begin{subequations}
\begin{eqnarray}
  ((\hat{G}^{(0)-1}-\underline{\hat{\Sigma}})*\underline{\hat{G}})(x_1;x_2)&=&\delta(x_1-x_2),
  \label{eq:Dyson1:G0}\\
  (\underline{\hat{G}}*(\hat{G}^{(0)-1}-\underline{\hat{\Sigma}}))(x_1;x_2)&=&\delta(x_1-x_2),
  \label{eq:Dyson2:G0}
\end{eqnarray}
\end{subequations}
where 
\begin{subequations}
\begin{eqnarray}
  \hat{G}^{(0)-1}(x_1;x_2) &=& 
  \left[\left(i\hbar D_{t_1}-\hat{H}_0(-i\hbar\mib{D}_{x_1})\right)\delta(x_1-x_2)\right]
  \\
  &=&\left[\left(-i\hbar D_{t_2}^*-\hat{H}_0(i\hbar\mib{D}_{x_2}^*)\right)\delta(x_1-x_2)\right].
\end{eqnarray}
\label{eq:G0^inv}%
\end{subequations}
It is known that the self-energy $\underline{\hat{\Sigma}}$ can be constructed from a Feynman rule~\cite{RammerSmith86} similar to that in the conventional Green's function method for $\hat{G}^{R,A}$ and $\hat{\Sigma}^{R,A}$.

\subsection{Wigner representation and non-commutativity as the Moyal product}
\label{subsec:non-commutativity}

We change the set of variables $(x_1;x_2)$ to $(X;x)$ with $X^\mu=(T,\mib{X})$ and $x^\mu=(t,\mib{x})$ being the center-of-mass and relative coordinates, respectively:
\begin{subequations}
\begin{eqnarray}
  X&\equiv& (x_1+x_2)/2,
  \label{eq:X}\\
  x&\equiv&  x_1-x_2.
  \label{eq:x}
\end{eqnarray}
\end{subequations}

We proceed to the Wigner representation in terms of $(X;p)$ by means of the Fourier transformation from $x^\mu$ to $p^\mu=({p_1}^\mu-{p_2}^\mu)/2=(\varepsilon,\mib{p})$ or from $x_\mu$ to $p_\mu=({p_1}_\mu-{p_2}_\mu)/2=(-\varepsilon,\mib{p})$. It is well known that the convolution given in Eq.~(\ref{eq:convolution:x_1;x_2}) can be rewritten in a compact form as a function of $X^\mu$ and $p_\mu$; 
\begin{equation}
  (\underline{\hat{\Upsilon}}*\underline{\hat{\Xi}})(X;p) = \underline{\hat{\Upsilon}}(X;p) e^{\frac{i\hbar}{2}(\overleftarrow{\partial}_{X^\mu} \overrightarrow{\partial}_{p_\mu}-\overleftarrow{\partial}_{p_\mu} \overrightarrow{\partial}_{X^\mu})} \underline{\hat{\Xi}}(X;p),
  \label{eq:convolution:X;p}
\end{equation}
which is explicitly proven in Appendix~\ref{app:Moyal}.
Henceforth, contraction of indices for four-component vectors should be understood. This takes the form of the so-called Moyal product or the star product of $\hat{\Upsilon}(X;p)$ and $\hat{\Xi}(X;p)$ in the space of $X^\mu$ and $p_\mu$, which fully represents the non-commutativity $[x^\mu_i,p^\mu_i]_-=i\hbar$. Recently, this has also been utilized in the field of high-energy physics in the context of the $D$-brane in string theory~\cite{SeibergWitten99}, and the associated non-commutative quantum mechanics~\cite{Non-commutativeQM} has been intensively investigated.

Now, in order to express the Dyson equations in the Wigner representation, we have only to calculate $\hat{G}^{(0)-1}(X;p)$. For this purpose, it is useful to define the quantity associated with the covariant derivative
\begin{equation}
  \pi_\mu(x_1;x_2)\equiv -i\hbar D_{{x_1}^\mu}\delta(x_1-x_2)
  =i\hbar D_{{x_2}^\mu}^*\delta(x_1-x_2),
  \label{eq:pi:x_1;x_2}
\end{equation}
so that $G^{(0)-1}(x_1;x_2)$ can be represented in the form of the convolutions of $\pi_\mu(x_1;x_2)$. The Wigner representation of Eq.~(\ref{eq:pi:x_1;x_2}) associated with the covariant derivative is obtained as
\begin{eqnarray}
  \pi_\mu(X;p)&=& \int\!dx\,e^{-ip_\mu x^\mu/\hbar}\pi_\mu\left(X+\frac{x}{2};X-\frac{x}{2}\right)
  \nonumber\\
  &=&p_\mu-qA_\mu(X)
  \label{eq:pi:X;p}
\end{eqnarray}
using integration by parts. This new variable $\pi^\mu(X;p)$ defines the mechanical or kinetic energy-momentum in the Wigner representation, and it is clearly distinct from the canonical one, $p^\mu$. 

Finally, the Wigner represetation of $\hat{G}^{(0)-1}$ can be obtained in a compact form by using the Moyal $*$ product as
\begin{equation}
  \hat{G}^{(0)-1}(X;p) = \hat{G}^{(0)-1}_{0*}(\pi(X;p)),
  \label{eq:G0^inv:X;p:def}
\end{equation}
where 
\begin{equation}
  \hat{G}^{(0)-1}_{0*}(\pi)=\pi^0-\hat{H}_{0*}(\mib{\pi})
  \label{eq:G0^inv_0*:pi}
\end{equation}
and $\hat{H}_{0*}(\pi)$ in the presence of electromagnetic fields are generalizations of 
\begin{equation}
  \hat{G}^{(0)-1}_0(p)=\varepsilon-\hat{H}_0(\mib{p})
  \label{eq:G0^inv_0:p}
\end{equation}
and $\hat{H}_0(p)$ defined in the absence of the fields. They are obtained by replacing all the products in $\hat{G}^{(0)-1}_0(p)$ and $\hat{H}_0(p)$ with the Moyal $*$ products defined by Eq.~(\ref{eq:convolution:X;p}) and also $p$ with $\pi=p-qA(X)$.
Therefore, in the Wigner representation, the Dyson equations (\ref{eq:Dyson1:G0}) and (\ref{eq:Dyson2:G0}) are rewritten as
\begin{subequations}
\begin{eqnarray}
  \left[\pi^0(X;p)-\hat{H}_{0*}(\mib{\pi}(X;p))-\underline{\hat{\Sigma}}(X;p)\right]*\underline{\hat{G}}(X;p)&=&1,
  \label{eq:Dyson1:X;p}\\
  \underline{\hat{G}}(X;p)*\left[\pi^0(X;p)-\hat{H}_{0*}(\mib{\pi}(X;p))-\underline{\hat{\Sigma}}(X;p)\right]&=&1.
  \label{eq:Dyson2:X;p}
\end{eqnarray}
\end{subequations}

\subsection{Gauge invariance in the case of constant electromagnetic fields}
\label{subsec:gauge}

Here, it is convenient to decompose the electromagnetic potential into two
parts; $\tilde{A}(X)$, which just contributes to uniform static
electromagnetic fields,
$\tilde{F}_{\mu\nu}=\partial_{x^\mu}\tilde{A}_\nu(x)\protect\linebreak-\partial_{x^\nu}\tilde{A}_\mu(x)={ 
{\rm constant}}$, and the other non-uniform and/or dynamical part, $\delta\!A(X)=A(X)-\tilde{A}(X)$. Similarly, we introduce 
\begin{equation}
  \tilde{\pi}^\mu(X;p)\equiv p^\mu-q\tilde{A}(X).
  \label{eq:pi^const}
\end{equation}
Now, we change the set of variables from $(X;p)$ to $(X|\tilde{\pi})$. Then, the Moyal product given by Eq.~(\ref{eq:convolution:X;p}) is transformed into the partly gauge-invariant form
\begin{equation}
  (\underline{\hat{\Upsilon}}*\underline{\hat{\Xi}})(X;p)=(\underline{\hat{\Upsilon}}\star\underline{\hat{\Xi}})(X|\tilde{\pi})=\underline{\hat{\Upsilon}}(X|\tilde{\pi})e^{\frac{i\hbar}{2}(\overleftarrow{\partial}_{X^\mu}\overrightarrow{\partial}_{\tilde{\pi}_\mu}-\overleftarrow{\partial}_{\tilde{\pi}_\mu}\overrightarrow{\partial}_{X^\mu}+q\tilde{F}^{\mu\nu}\overleftarrow{\partial}_{\tilde{\pi}^\mu}\overrightarrow{\partial}_{\tilde{\pi}^\nu})}\underline{\hat{\Xi}}(X|\tilde{\pi}).
  \label{eq:convolution:X|pi^const}
\end{equation}
In this $(X|\tilde{\pi})$ representation, the Dyson equations take the forms
\begin{subequations}
\begin{eqnarray}
  \left[\tilde{\pi}^0-q\delta\!\phi(X)-\hat{H}_{0\star}(\tilde{\mib{\pi}}-q\,\delta\!\mib{A}(X))-\underline{\hat{\Sigma}}(X|\tilde{\pi})\right]\star\underline{\hat{G}}(X|\tilde{\pi})&=&1,
  \label{eq:Dyson1:X|pi^const}\\
  \underline{\hat{G}}(X|\tilde{\pi})\star\left[\tilde{\pi}^0-q\delta\!\phi(X)-\hat{H}_{0\star}(\tilde{\mib{\pi}}-q\,\delta\!\mib{A}(X))-\underline{\hat{\Sigma}}(X|\tilde{\pi})\right]&=&1.
  \label{eq:Dyson2:X|pi^const}
\end{eqnarray}
\end{subequations}

In the rest of this paper, we restrict ourselves to the case of a uniform
static applied electromagnetic field $F^{\mu\nu}=\partial_{x_\mu}
A^\nu-\partial_{x_\nu} A^\mu={\rm constant}$, i.e., $A(X)=\tilde{A}(X)$, or equivalently, $\delta\!A(x)=0$. Now we do not have to distinguish between $\pi$ and $\tilde{\pi}$. In the case of the constant electromagnetic fields, the Moyal $\star$ product in the $(X|\pi)$ space, which is defined by Eq.~(\ref{eq:convolution:X|pi^const}), gives the gauge-invariant representation of the conventional Moyal $*$ product given by Eq.~(\ref{eq:convolution:X;p}). Namely, like $\hat{\Upsilon}(X|\pi)$ and $\hat{\Xi}(X|\pi)$, their convolution has only an implicit dependence on the choice of the gauge through $\pi^\mu=p^\mu-eA^\mu(X)$, and hence the gauge invariance after the $p$ or $\pi$ integration is evident. This is consistent with Schwinger's idea to extract gauge-invariant results by using gauge-covariant quantities~\cite{Schwinger51,LevandaFleurov94,Kita01}, since $\pi^\mu(X;p)$ is constructed from the gauge-covariant derivative via Eqs.~(\ref{eq:pi:x_1;x_2}) and (\ref{eq:pi:X;p}). Furthermore, in the present case of constant electromagnetic fields, imposing the fully symmetric combination of the Moyal products among $\pi$ for the Hamiltonian $\hat{H}_{0*}(\mib{\pi})$, as required for quantum-mechanical problems, the $\star$ products implicitly appearing in $\hat{H}_0(\mib{\pi})$ are reduced to conventional products, owing to the antisymmetric property $F_{\mu\nu}+F_{\nu\mu}=0$. Therefore, we arrive at the following gauge-invariant forms of the Dyson equations under a constant electromangeitc field:
\begin{subequations}
\begin{eqnarray}
  \left[\pi^0-\hat{H}_0(\mib{\pi})-\underline{\hat{\Sigma}}(X|\pi)\right]\star\underline{\hat{G}}(X|\pi)&=&1,
  \label{eq:Dyson1:X|pi:const}\\
  \underline{\hat{G}}(X|\pi)\star\left[\pi^0-\hat{H}_0(\mib{\pi})-\underline{\hat{\Sigma}}(X|\pi)\right]&=&1.
  \label{eq:Dyson2:X|pi:const}
\end{eqnarray}
\end{subequations}

Furthermore, as long as we restrict ourselves to a uniform, stationary solution, Eq.~(\ref{eq:convolution:X|pi^const}) reduces to
\begin{equation}
  (\underline{\hat{\Upsilon}}\star\underline{\hat{\Xi}})(\pi) = \underline{\hat{\Upsilon}}(\pi) e^{\frac{i\hbar q}{2}F^{\mu\nu}\overleftarrow{\partial}_{\pi^\mu}\overrightarrow{\partial}_{\pi^\nu}}\underline{\hat{\Xi}}(\pi).
  \label{eq:convolution:X|pi_0_const.field_dX=0}
\end{equation}
This, again, takes the form of the Moyal product, but now it translates the non-commutativity among the components of the mechanical energy-momentum oprator ${\pi_i}^\mu={p_i}^\mu-eA^\mu(x_i)$, which is introduced by the physical electromagnetic field $\hbar qF^{\mu\nu}\ne0$. Then, using this Moyal product defined by Eq.~(\ref{eq:convolution:X|pi_0_const.field_dX=0}), the Dyson equations~(\ref{eq:Dyson1:X|pi:const}) and (\ref{eq:Dyson2:X|pi:const}) can be expressed in the gauge-invariant forms
\begin{subequations}
\begin{eqnarray}
  (\pi^0-\hat{H}_0(\mib{\pi})-\underline{\hat{\Sigma}}(\pi))\star\underline{\hat{G}}(\pi)&=&1,
  \label{eq:Dyson1:Moyal}\\
  \underline{\hat{G}}(\pi)\star(\pi^0-\hat{H}_0(\mib{\pi})-\underline{\hat{\Sigma}}(\pi))&=&1.
  \label{eq:Dyson2:Moyal}
\end{eqnarray}
\label{eq:Dyson:Moyal}%
\end{subequations}
These two Dyson equations guarantee that for a given $\underline{\hat{\Sigma}}$, the solution for $\underline{\hat{G}}$ is unique. Therefore, one can define the inverse with respect to the Moyal product as
\begin{equation}
  \underline{\hat{G}}(\pi)=\left(\pi^0-\hat{H}_0(\mib{\pi})-\underline{\hat{\Sigma}}(\pi)\right)^{-1}_{\star}.
    \label{eq:_G_:inverse}
\end{equation}

\subsection{Self-energy due to the scattering by dilute impurities}
\label{subsec:impurity}

Here, we consider the self-energy arising only from the scattering by dilute impurities. For a general impurity potential of the form
\begin{equation}
  \hat{V}(\mib{x})=\sum_i\hat{u}(\mib{x}-\mib{x}_i),
  \label{eq:imp}
\end{equation}
the simplest self-consistent Born approximation reads
\begin{eqnarray}
  \underline{\hat{\Sigma}}(\pi^0,\mib{\pi}) &=& n_{\rm imp}\int\!\frac{d\mib{\pi}}{(2\pi\hbar)^d}\hat{u}_{\mib{\pi}-\mib{\pi}'}\underline{\hat{G}}(\pi^0,\mib{\pi}')\hat{u}_{\mib{\pi}'-\mib{\pi}},
  \label{eq:_Sigma_:Born_1}\\
  \hat{u}_{\mib{q}}&=&\int\!d\mib{x}e^{-i\mib{q\cdot x}}\hat{u}(\mib{x}).
  \label{eq:u_q}
\end{eqnarray}
When the potential is given by a $\delta$-function, i.e., 
\begin{subequations}
\begin{eqnarray}
  \hat{u}(\mib{x})&=&\hat{u}\delta(\mib{x}),
  \label{eq:u_x:delta}\\
  \hat{u}_{\mib{q}}&=&\hat{u}, 
  \label{eq:u_q:delta}
\end{eqnarray}
\label{eq:u:delta}%
\end{subequations}
it is easy to construct the self-consistent equations in terms of the $T$-matrix approximation, taking account of an inifinite series consisting of the first and all the multiple Born scattering amplitudes for the self-energy $\underline{\hat{\Sigma}}$, as
\begin{eqnarray}
  \underline{\hat{\Sigma}}(\pi^0) &=& n_{\rm imp}\underline{\hat{T}}(\pi^0),
  \label{eq:_Sigma_:Born_all}\\
  \underline{\hat{T}}(\pi^0) &=& \hat{u}\left(1-\underline{\hat{g}}(\pi^0)\hat{u}\right)^{-1},
  \label{eq:_T_}\\
  \underline{\hat{g}}(\pi^0) &=& \int\!\frac{d\mib{\pi}}{(2\pi\hbar)^d}\underline{\hat{G}}(\pi^0,\mib{\pi}).
  \label{eq:_g_}
\end{eqnarray}

\subsection{Expansion in the electromagnetic fields}
\label{subsec:exp}

In most cases, except for the simplest problems, like the Zener problem~\cite{Zener}, the Dyson equations~(\ref{eq:Dyson1:Moyal}) and (\ref{eq:Dyson2:Moyal}) cannot be directly solved in the context of non-commutative quantum mechanics using the Moyal product. Therefore, the perturbative treatment is an important and powerful tool.

Dyson equations in the form of Eqs.~(\ref{eq:Dyson:Moyal}) can be expanded in $\hbar qF^{\mu\nu}$ as
\begin{subequations}
\begin{eqnarray}
  &&\left(\pi^0-\hat{H}_0(\mib{\pi})-\underline{\hat{\Sigma}}(\pi)\right)\underline{\hat{G}}(\pi)-1
 \nonumber \\
  &=&-\sum_{n=1}^\infty\frac{1}{n!}\left(\pi^0-\hat{H}_0(\mib{\pi})-\underline{\hat{\Sigma}}(\pi)\right)\left(\prod_{i=1}^n\frac{i\hbar q F^{\mu_i\nu_i}}{2}\overleftarrow{\partial}_{\pi^{\mu_i}}\overrightarrow{\partial}_{\pi^{\nu_i}}\right)\underline{\hat{G}}(\pi),
   \label{eq:Dyson1:grad}\nonumber\\
\\
 &&\underline{\hat{G}}(\pi)\left(\pi^0-\hat{H}_0(\mib{\pi})-\underline{\hat{\Sigma}}(\pi)\right)-1
 \nonumber\\
 &=&-\sum_{n=1}^\infty\frac{1}{n!}\underline{\hat{G}}(\pi)\left(\prod_{i=1}^n\frac{i\hbar q F^{\mu_i\nu_i}}{2}\overleftarrow{\partial}_{\pi^{\mu_i}}\overrightarrow{\partial}_{\pi^{\nu_i}}\right)\left(\pi^0-\hat{H}_0(\mib{\pi})-\underline{\hat{\Sigma}}(\pi)\right).
 \label{eq:Dyson2:grad}\nonumber\\
\end{eqnarray}
\label{eq:_Dyson_:grad}%
\end{subequations}
Finally, we obtain
\begin{subequations}
\begin{eqnarray}
  \underline{\hat{G}}(\pi)&=&\underline{\hat{G}}_0(\pi)\left[1+\left(\underline{\hat{\Sigma}}(\pi)-\underline{\hat{\Sigma}}_0(\pi)\right)\underline{\hat{G}}(\pi)\right.
    \nonumber\\
    &&\left.-\sum_{n=1}^\infty\frac{1}{n!}\left(\pi^0-\hat{H}_0(\mib{\pi})-\underline{\hat{\Sigma}}(\pi)\right)\left(\prod_{i=1}^n\frac{i\hbar qF^{\mu_i\nu_i}}{2}\overleftarrow{\partial}_{\pi^{\mu_i}}\overrightarrow{\partial}_{\pi^{\nu_i}}\right)\underline{\hat{G}}(\pi)\right]\\
  \label{eq:_G_1:grad}
  &=&\left[1+\underline{\hat{G}}(\pi)\left(\underline{\hat{\Sigma}}(\pi)-\underline{\hat{\Sigma}}_0(\pi)\right)\right.
    \nonumber\\
    &&\left.\ {}-\sum_{n=1}^\infty\frac{1}{n!}\underline{\hat{G}}(\pi)\left(\prod_{i=1}^n\frac{i\hbar qF^{\mu_i\nu_i}}{2}\overleftarrow{\partial}_{\pi^{\mu_i}}\overrightarrow{\partial}_{\pi^{\nu_i}}\right)\left(\pi^0-\hat{H}_0(\mib{\pi})-\underline{\hat{\Sigma}}(\pi)\right)\right]\underline{\hat{G}}_0(\pi).
  \nonumber\\
  \label{eq:_G_2:grad}
\end{eqnarray}
\label{eq:_G_:grad}%
\end{subequations}
Henceforth, quantities with the subscript ``$0$'' denote those defined in the absence of external electromagnetic fields. This notation is introduced in order to distinguish them from those in the presence of the fields. 

We note that a non-linear generalization of the quantum Boltzmann equation in the steady, uniform case can be derived from Eqs.~(\ref{eq:_Dyson_:grad}) as
\begin{eqnarray}
  &&\left[\underline{\hat{G}}(\pi),\hat{H}_0(\mib{\pi})+\underline{\hat{\Sigma}}(\pi)\right]_-
  \nonumber\\
  &=&\sum_{n=1}^\infty\frac{1}{n!}\left[\underline{\hat{G}}(\pi)\left(\prod_{i=1}^n\frac{i\hbar q F^{\mu_i\nu_i}}{2}\overleftarrow{\partial}_{\pi^{\mu_i}}\overrightarrow{\partial}_{\pi^{\nu_i}}\right)\left(\pi^0-\hat{H}_0(\mib{\pi})-\underline{\hat{\Sigma}}(\pi)\right)
    \right.\nonumber\\
  &&\left.\ \ \ \ \ \ {}-\left(\pi^0-\hat{H}_0(\mib{\pi})-\underline{\hat{\Sigma}}(\pi)\right)\left(\prod_{i=1}^n\frac{i\hbar q F^{\mu_i\nu_i}}{2}\overleftarrow{\partial}_{\pi^{\mu_i}}\overrightarrow{\partial}_{\pi^{\nu_i}}\right)\underline{\hat{G}}(\pi)\right]
  \label{eq:_QBE_:total}
\end{eqnarray}
in the Keldysh space. It is clear that this quantum Boltzmann equation cannot determine the retarded and advanced components $\hat{G}^{R,A}(\pi)$ and $\hat{\Sigma}^{R,A}(\pi)$, as readily verified for single-component, i.e., scalar, cases. Furthermore, in multi-component systems, it does not uniquely determine $\hat{G}^<(\pi)$ and $\hat{\Sigma}^<(\pi)$ either, as we will show later.

Equations~(\ref{eq:_G_1:grad}) and (\ref{eq:_G_2:grad}) have the solutions of the forms
\begin{eqnarray}
  \underline{\hat{G}}(\pi)&=&\underline{\hat{G}}_0(\pi)+\sum_{n=1}^\infty \frac{1}{n!}\left(\prod_{i=1}^n \frac{\hbar qF^{\mu_i\nu_i}}{2}\right)\underline{\hat{G}}_{\mu_1\nu_1,\cdots,\mu_n\nu_n}(\pi),
\label{eq:_G_:exp}\\
  \underline{\hat{\Sigma}}(\pi)&=&\underline{\hat{\Sigma}}_0(\pi)+\sum_{n=1}^\infty \frac{1}{n!}\left(\prod_{i=1}^n \frac{\hbar qF^{\mu_i\nu_i}}{2}\right)\underline{\hat{\Sigma}}_{\mu_1\nu_1,\cdots,\mu_n\nu_n}(\pi).
\label{eq:_Sigma_:exp}
\end{eqnarray}
We can calculate $\underline{\hat{G}}(\pi)$ and $\underline{\hat{\Sigma}}(\pi)$ by equating the coefficients of each order in $F^{\mu\nu}$ in Eq.~(\ref{eq:_G_1:grad}) or (\ref{eq:_G_2:grad}) and, for instance, imposing that $\underline{\hat{G}}_{\mu_1\nu_1,\cdots,\mu_n\nu_n}$ and $\underline{\hat{\Sigma}}_{\mu_1\nu_1,\cdots,\mu_n\nu_n}$ are symmetric under the permutation $(\mu_i\nu_i)\leftrightarrow(\mu_j\nu_j)$ and anti-symmetric under the permutation $\mu_i\leftrightarrow\nu_i$.

In particular, in the case of impurity problems, we obtain
\begin{equation}
  \underline{\hat{\Sigma}}_{\mu_1\nu_1,\cdots,\mu_n\nu_n}(\pi^0,\mib{\pi})=\int\!\frac{d\mib{\pi}'}{(2\pi\hbar)^d}\hat{u}_{\mib{\pi}-\mib{\pi}'}\underline{\hat{G}}_{\mu_1\nu_1,\cdots,\mu_n\nu_n}(\pi^0,\mib{\pi}')\hat{u}_{\mib{\pi}'-\mib{\pi}}
  \label{eq:_Sigma_:Born_1:exp}
\end{equation}
in the self-consistent Born approximation for the general impurity pontential given in Eq.~(\ref{eq:u_q}), and 
\begin{eqnarray}
  \underline{\hat{\Sigma}}_{\mu_1\nu_1,\cdots,\mu_n\nu_n}(\pi^0)&=&n_{\text{imp}}\underline{\hat{T}}_0(\pi^0)\sum_{\frak{s}}\sum_{i=1}^n\mathop{{\sum}'}_{m_1,\cdots,m_{i-1}}\prod_{j=1}^i
  \nonumber\\
  &&\hspace*{-90pt}\times\left[\frac{1}{(m_j-m_{j-1})!} \ \underline{\hat{g}}_{\mu_{\frak{s}(m_{j-1}+1)}\nu_{\frak{s}(m_{j-1}+1)},\cdots,\mu_{\frak{s}(m_j)}\nu_{\frak{s}(m_j)}}(\pi^0) \ \underline{\hat{T}}_0(\pi^0)\right],
 % \nonumber\\
  \label{eq:_Sigma_:Born_all:exp}\\
  \underline{\hat{g}}_{\mu_1\nu_1,\cdots,\mu_n\nu_n}(\pi^0)&=&\int\!\frac{d\mib{\pi}}{(2\pi\hbar)^d}\underline{\hat{G}}_{\mu_1\nu_1,\cdots,\mu_n\nu_n}(\pi^0,\mib{\pi})
  \label{eq:_g_:exp}
\end{eqnarray}
for the $T$-matrix approximation with the $\delta$-function impurity potential given in Eq.~(\ref{eq:u:delta}), where $m(0)=0$ and $m(i)=n$, and $\sum_{m(1),\cdots,m(i-1)}^\prime$ and $\sum_{\frak{s}}$ denote the summations over $m(1),\cdots,m(i-1)$ under the condition $m(j)<m(j')$ for $j<j'$ and over all the permutations among $1,\cdots,n$, respectively.

\subsection{Electric current}
\label{subsec:current}

Here, we summarize the formulation of the electric currents in the presence of constant electromagnetic fields.

The component-dependent particle number and current density at a given time-space coordinate $X^\mu$, $\hat{\rho}^\mu(X)=(\hat{n}(X),\hat{\mib{j}}(X))$, are expressed in terms of the quantum distribution function $\hat{G}^<$ as
\begin{eqnarray}
  \hat{n}(X) &=&\left. -i\hat{G}^<\left(X+\frac{x}{2};X-\frac{x}{2}\right)\right|_{x\to0}
  =\int_{-\infty}^\infty\!\frac{d\pi^0}{2\pi i}\int\!\frac{d^d\mib{\pi}}{(2\pi\hbar)^d}\hat{G}^<(X|\pi),
  \label{eq:den}\\
  \hat{\mib{j}}(X)
  &=&\int_{-\infty}^\infty\!\frac{d\pi^0}{2\pi i}\int\!\frac{d^d\mib{\pi}}{(2\pi\hbar)^d}\frac{1}{2}\left[\hat{\mib{v}}_0(\mib{\pi}),\hat{G}^<(X|\pi)\right]_+,
\end{eqnarray}
with the velocity 
\begin{equation}
  \hat{\mib{v}}(\mib{p})=\frac{1}{i\hbar}\left[\mib{x},\hat{H}^\text{tot}(x,\mib{p})\right]_-
  =\mib{\nabla}_p\hat{H}(x,\mib{p})
  =\mib{\nabla}_\pi\hat{H}_0(\mib{\pi})=\hat{\mib{v}}_0(\mib{\pi})
  \label{eq:v}
\end{equation}
which satisfies the desired condition of the equation of continuity for the total particle number,
\begin{equation}
  \partial_{X^\mu}\rho^\mu(X)=0,
  \label{eq:eq-continuity}
\end{equation}
with the total number and current density $\rho^\mu(X)\equiv \text{Tr}\hat{\rho}^\mu(X)$.

Finally, the uniform, static electric currents are given by the average over both time and space
\begin{equation}
  \mib{J} = q\lim_{T\to\infty}\lim_{V\to\infty}\frac{1}{TV}\int_{-T/2}^{T/2}\!dT'\,\int_V\!d\mib{X}'\,{\rm Tr}\left[\hat{\mib{j}}(X')\right].
  \label{eq:J}
\end{equation}
It is easy to see that $\hat{G}^<_0(\pi)$ never contributes to the current $\mib{J}$ without breaking the inversion symmetry in the absence of an electric field.

\subsection{Zeroth-order equilibrium properties}
\label{subsec:equilibrium}

Lastly, for later use, we summarize the well-known equilibrium properties in the absence of external electromagnetic fields:
\begin{eqnarray}
  \hat{G}^{R(A)}_0(p) &=& \left(\varepsilon-\hat{H}_0(\mib{p})-\hat{\Sigma}^{R(A)}_0(p)\right)^{-1},
  \label{eq:G^R(A)_0}\\
  \hat{G}^<_0(p) &=& \left(\hat{G}^A_0(p)- \hat{G}^R_0(p)\right)f_\mp(\varepsilon),
  \label{eq:G^<_0}\\
  \hat{\Sigma}^<_0(p) &=& \left(\hat{\Sigma}^A_0(p)-\hat{\Sigma}^R_0(p)\right)f_\mp(\varepsilon),
  \label{eq:Sigma^<_0}
\end{eqnarray}
with the Bose and the Fermi distribution functions $f_-(\varepsilon)=(1-\coth(\beta(\varepsilon-\mu)/2))/2$ and $f_+(\varepsilon)=(1-\tanh(\beta(\varepsilon-\mu)/2))/2$, respectively.

The self-energy is obtained as 
\begin{equation}
  \hat{\Sigma}^{R(A)}_0(\varepsilon) = n_{\rm imp}\int\!\frac{d\mib{p}}{(2\pi\hbar)^d}\hat{u}_{\mib{p}-\mib{p}'}\hat{G}^{R(A)}_0(\varepsilon,\mib{p}')\hat{u}_{\mib{p}'-\mib{p}}
  \label{eq:Sigma^R,A:0:Born_1}
\end{equation}
in the Born approximation for the impurity potential given by Eq.~(\ref{eq:u_q}), and as
\begin{eqnarray}
  \hat{\Sigma}^{R(A)}_0(\varepsilon) &=& n_{\rm imp}\hat{T}^{R(A)}_0(\varepsilon),
  \label{eq:Sigma^R,A:0:Born_all}\\
  \hat{T}^{R(A)}_0(\varepsilon) &=& \hat{u}\left(1-\hat{g}^{R(A)}_0(\varepsilon)\hat{u}\right)^{-1},
  \label{eq:T^R,A:0:Born_all}\\
  \hat{g}^{R(A)}_0(\varepsilon) &=& \int\!\frac{d\mib{p}}{(2\pi\hbar)^d}\hat{G}^{R(A)}_0(\varepsilon,\mib{p})
  \label{eq:g^R,A:0}
\end{eqnarray}
in the $T$-matrix approximation for the $\delta$-function potential given by Eq.~(\ref{eq:u:delta}).

\section{Linear response theory}
\label{sec:1st}

We proceed to the linear response of the Green's functions to the external fields $F_{\mu\nu}$. In particular, we must calculate $\hat{G}^<_{\mu\nu}$ to consider the linear response of various quantities. For this purpose, we linearize Eq.~(\ref{eq:_G_1:grad}) or (\ref{eq:_G_2:grad}) in $F_{\mu\nu}$ to obtain
\begin{equation}
  \underline{\hat{G}}_{\mu\nu}(\pi) = \underline{\hat{G}}_0(\pi)\left[\underline{\hat{\Sigma}}_{\mu\nu}(\pi)+\frac{i}{2}\left((\partial_{\pi^\mu}\underline{\hat{G}}_0^{-1}(\pi))\underline{\hat{G}}_0(\pi)(\partial_{\pi^\nu}\underline{\hat{G}}_0^{-1}(\pi))-(\mu\leftrightarrow\nu)\right)\right]\underline{\hat{G}}_0(\pi),
  \label{eq:_G_:grad1}
\end{equation}
or 
\begin{eqnarray}
  \underline{\hat{G}}_{\mib{E}}(\pi) &=& \underline{\hat{G}}_0(\pi)\left[\underline{\hat{\Sigma}}_{\mib{E}}(\pi)+\frac{i}{2}\left((\partial_{\pi^0}\underline{\hat{G}}_0^{-1}(\pi))\underline{\hat{G}}_0(\pi)(\mib{\nabla}_\pi\underline{\hat{G}}_0^{-1}(\pi))
    \right.\right.\nonumber\\
    &&\left.\left.\ \ \ \ \ \ \ \ \ \ \ \ \ \ \ \ \ \ \ \ \ \ {}-(\mib{\nabla}_\pi\underline{\hat{G}}_0^{-1}(\pi))\underline{\hat{G}}_0(\pi)(\partial_{\pi^0}\underline{\hat{G}}_0^{-1}(\pi))\right)\right]\underline{\hat{G}}_0(\pi),
  \label{eq:_G_:E}\\
  \underline{\hat{G}}_{\mib{B}}(\pi) &=& \underline{\hat{G}}_0(\pi)\left[\underline{\hat{\Sigma}}_{\mib{B}}(\pi)+\frac{i}{2}(\mib{\nabla}_\pi\underline{\hat{G}}_0^{-1}(\pi))\times\underline{\hat{G}}_0(\pi)(\mib{\nabla}_\pi\underline{\hat{G}}_0^{-1}(\pi))\right]\underline{\hat{G}}_0(\pi).
  \label{eq:_G_:B}
\end{eqnarray}
Here, we have used the identity
\begin{equation}
  \partial_{\pi^\mu}\underline{\hat{G}}_0(\pi)=-\underline{\hat{G}}_0(\pi)\left(\partial_{\pi^\mu}\underline{\hat{G}}_0^{-1}(\pi)\right)\underline{\hat{G}}_0(\pi),
  \label{eq:diff}
\end{equation}
or equivalently
\begin{subequations}
\begin{eqnarray}
  \partial_{\pi^\mu}\hat{G}_0^{R(A)}(\pi)
  &=&{}-\hat{G}^{R(A)}_0(\pi)\left(\partial_{\pi^\mu}\hat{G}^{R(A) -1}_0(\pi))\right)\hat{G}_0^{R(A)}(\pi),
  \label{eq:diff:R,A}\\
  \partial_{\pi^\mu}\hat{G}_0^<(\pi)
  &=&{}-\left[\hat{G}^A_0(\pi)\left(\partial_{\pi^\mu}\hat{G}^{A -1}_0(\pi)\right)\hat{G}_0^A(\pi)
    \right.
    \nonumber\\
    &&\left.\ {}-\hat{G}^R_0(\pi)\left(\partial_{\pi^\mu}\hat{G}^{R -1}_0(\pi)\right)\hat{G}_0^R(\pi)\right]f_\mp(\pi^0)
  \nonumber\\
  &&{}+(\hat{G}^A_0(\pi)-\hat{G}^R_0(\pi))\partial_{\pi^\mu}f_\mp(\pi^0).
  \label{eq:diff:<}
\end{eqnarray}
\end{subequations}

\subsection{Retarded and advanced Green's functions and self-energies}
\label{subsec:1st:R,A}

First, we consider the retarded (advanced) component of the Dyson equation~(\ref{eq:_G_:grad1}), which reads
\begin{eqnarray}
  \hat{G}^{R(A)}_{\mib{E}}(\pi)&=&\hat{G}^{R(A)}_0(\pi)\left[
    \hat{\Sigma}^{R(A)}_{\mib{E}}(\pi)
    \right.\nonumber\\
    &&\ \ \ \ \ \ \ \ \ \ {}+\frac{i}{2}
    \left((\partial_{\pi^0}\hat{G}^{R(A)-1}_0(\pi))\hat{G}^{R(A)}_0(\pi)(\mib{\nabla}_\pi\hat{G}^{R(A)-1}_0(\pi))
    \right.\nonumber\\
    &&\left.\left.\ \ \ \ \ \ \ \ \ \ \ \ \ \ {}-(\mib{\nabla}_\pi\hat{G}^{R(A)-1}_0(\pi))\hat{G}^{R(A)}_0(\pi)(\partial_{\pi^0}\hat{G}^{R(A)-1}_0(\pi))\right)\right]\hat{G}^{R(A)}_0(\pi),
  \nonumber\\
  \label{eq:G^R,A:E}\\
  \hat{G}^{R(A)}_{\mib{B}}(\pi)&=&\hat{G}^{R(A)}_0(\pi)\left[
    \hat{\Sigma}^{R(A)}_{\mib{B}}(\pi)
    \right.\nonumber\\
    &&\left.\ \ \ \ \ \ \ \ \ \ {}+\frac{i}{2}(\mib{\nabla}_\pi\hat{G}^{R(A)-1}_0(\pi))\times\hat{G}^{R(A)}_0(\pi)(\mib{\nabla}_\pi\hat{G}^{R(A)-1}_0(\pi))\right]\hat{G}^{R(A)}_0(\pi),
  \nonumber\\
  \label{eq:G^R,A:B}%
\end{eqnarray}
because of the relations given by Eqs.~(\ref{eq:app:U^R}) and (\ref{eq:app:U^A}).

The self-energies due to the impurity scattetring are obtained from Eqs.~(\ref{eq:_Sigma_:Born_1}) and (\ref{eq:_Sigma_:Born_all}) as
\begin{eqnarray}
  \hat{\Sigma}^{R(A)}_{\mib{E}}(\pi^0,\mib{\pi})&=&n_{\text{imp}}\int\!\frac{d\mib{\pi}'}{(2\pi\hbar)^d}\hat{u}_{\mib{\pi}-\mib{\pi}'}\hat{G}^{R(A)}_{\mib{E}}(\pi^0,\mib{\pi}')\hat{u}_{\mib{\pi}'-\mib{\pi}},
  \label{eq:Sigma^R,A:E:Born_1}\\
  \hat{\Sigma}^{R(A)}_{\mib{B}}(\pi^0,\mib{\pi})&=&n_{\text{imp}}\int\!\frac{d\mib{\pi}'}{(2\pi\hbar)^d}\hat{u}_{\mib{\pi}-\mib{\pi}'}\hat{G}^{R(A)}_{\mib{B}}(\pi^0,\mib{\pi}')\hat{u}_{\mib{\pi}'-\mib{\pi}},
  \label{eq:Sigma^R,A:B:Born_1}
\end{eqnarray}
in the Born approximation for the general potential, Eq.~(\ref{eq:u_q}), and as
\begin{eqnarray}
  \hat{\Sigma}^{R(A)}_{\mib{E}}(\pi^0)&=&n_{\text{imp}}\hat{T}^{R(A)}_0(\pi^0)\hat{g}^{R(A)}_{\mib{E}}(\pi^0)\hat{T}^{R(A)}_0(\pi^0),
  \label{eq:Sigma^R,A:E:Born_all}\\
  \hat{\Sigma}^{R(A)}_{\mib{B}}(\pi^0)&=&n_{\text{imp}}\hat{T}^{R(A)}_0(\pi^0)\hat{g}^{R(A)}_{\mib{B}}(\pi^0)\hat{T}^{R(A)}_0(\pi^0),
  \label{eq:Sigma^R,A:B:Born_all}
\end{eqnarray}
in the $T$-matrix approximation for the $\delta$-function potential, Eq.~(\ref{eq:u:delta}).

\subsection{Lesser Green's functions}
\label{subsec:1st:<}

Using the relation given by Eq.~(\ref{eq:app:U^<}), the lesser component of Eqs.~(\ref{eq:_G_:E}) and (\ref{eq:_G_:B}) can be expressed as
\begin{eqnarray}
  \hat{G}^<_{\mib{E}}(\pi)
  &=&\left(\hat{G}^A_{\mib{E}}(\pi)-\hat{G}^R_{\mib{E}}(\pi)\right)f_\mp(\pi^0)
  \nonumber\\
  &&{}+\hat{G}^R_0(\pi)\left(\hat{\Sigma}^<_{\mib{E}}-(\hat{\Sigma}^A_{\mib{E}}(\pi)-\hat{\Sigma}^R_{\mib{E}}(\pi))f_\mp(\pi^0)\right)\hat{G}^A_0(\pi)
  \nonumber\\
  &&{}-\frac{i}{2}\hat{G}^R_0(\pi)\left((\hat{\Sigma}^A_0(\pi)-\hat{\Sigma}^R_0(\pi))\hat{G}^A_0(\pi)(\mib{\nabla}_\pi\hat{G}^{A-1}_0(\pi))\right.
  \nonumber\\
  &&\ \ \ \left.{}-(\mib{\nabla}_\pi\hat{G}^{R-1}_0(\pi))\hat{G}^R_0(\pi)(\hat{\Sigma}^A_0(\pi)-\hat{\Sigma}^R_0(\pi))\right)\hat{G}^A_0(\pi)\partial_{\pi^0}f_\mp(\pi^0)
\ \ \ \ \ 
  \label{eq:G^<:E}
\end{eqnarray}
and
\begin{eqnarray}
  \hat{G}^<_{\mib{B}}(\pi)
  &=&\left(\hat{G}^A_{\mib{B}}(\pi)-\hat{G}^R_{\mib{B}}(\pi)\right)f_\mp(\pi^0)
  \nonumber\\
  &&{}+\hat{G}^R_0(\pi)\left(\hat{\Sigma}^<_{\mib{B}}-(\hat{\Sigma}^A_{\mib{B}}(\pi)-\hat{\Sigma}^R_{\mib{B}}(\pi))f_\mp(\pi^0)\right)\hat{G}^A_0(\pi).
  \label{eq:G^<:B}%
\end{eqnarray}
These sets of equations have the solutions of the form
\begin{subequations}
\begin{eqnarray}
  \hat{G}^<_{\mib{E}}(\pi) &=& \hat{G}^<_{\mib{E},I}(\pi)\partial_{\pi^0} f_\mp(\pi^0)+\hat{G}^<_{\mib{E},II}(\pi)f_\mp(\pi^0),
  \label{eq:G^<:E:I,II}\\
  \hat{\Sigma}^<_{\mib{E}}(\pi) &=& \hat{\Sigma}^<_{\mib{E},I}(\pi)\partial_{\pi^0} f_\mp(\pi^0)+\hat{\Sigma}^<_{\mib{E},II}(\pi)f_\mp(\pi^0),
  \label{eq:Sigma^<:E:I,II}
\end{eqnarray}
\label{eq:G,Sigma^<:E:I,II}%
\end{subequations}
and
\begin{subequations}
\begin{eqnarray}
  \hat{G}^<_{\mib{B}}(\pi) &=& \hat{G}^<_{\mib{B},I}(\pi)\partial_{\pi^0} f_\mp(\pi^0)+\hat{G}^<_{\mib{B},II}(\pi)f_\mp(\pi^0),
  \label{eq:G^<:B:I,II}\\
  \hat{\Sigma}^<_{\mib{B}}(\pi) &=& \hat{\Sigma}^<_{\mib{B},I}(\pi)\partial_{\pi^0} f_\mp(\pi^0)+\hat{\Sigma}^<_{\mib{B},II}(\pi)f_\mp(\pi^0),
  \label{eq:Sigma^<:B:I,II}
\end{eqnarray}
\label{eq:G,Sigma^<:B:I,II}%
\end{subequations}
respectively.
The first and the second terms represent the classical contributions from the states at the chemical potential with a factor of $\partial_{\pi^0}f_\mp(\pi^0)$ and the quantum contributions from all the occupied states with $f_\mp(\pi^0)$, respectively. Substituting Eqs.~(\ref{eq:G^<:E:I,II}) and (\ref{eq:Sigma^<:E:I,II}) into Eq.~(\ref{eq:G^<:E}) and Eqs.~(\ref{eq:G^<:B:I,II}) and (\ref{eq:Sigma^<:B:I,II}) into Eq.~(\ref{eq:G^<:B}), we finally obtain
\begin{eqnarray}
  \hat{G}^<_{\mib{E},I}(\pi)&=&\hat{G}^R_0(\pi)\hat{\Sigma}^<_{\mib{E},I}(\pi)\hat{G}^A_0(\pi)
  \nonumber\\
  &&{}-\frac{i}{2}\left(\hat{G}^R_0(\pi)(\mib{\nabla}_\pi(\hat{H}_0(\mib{\pi})+\hat{\Sigma}^R_0(\pi)))(\hat{G}^A_0(\pi)-\hat{G}^R_0(\pi))\right.
  \nonumber\\
  &&\left.\ \ \ \ \ {}-(\hat{G}^A_0(\pi)-\hat{G}^R_0(\pi))(\mib{\nabla}_\pi(\hat{H}_0(\mib{\pi})+\hat{\Sigma}^A_0(\pi)))\hat{G}^A_0(\pi)\right)
  \nonumber\\
  &=&\hat{G}^R_0(\pi)\left(\hat{\Sigma}^<_{\mib{E},I}(\pi)-i\mib{\nabla}_\pi\left(\hat{H}_0(\mib{\pi})+\frac{1}{2}(\hat{\Sigma}^R_0(\pi)+\hat{\Sigma}^A_0(\pi))\right)\right)\hat{G}^A_0(\pi)
  \nonumber\\
  &&+\frac{i}{2}\mib{\nabla}_\pi\left(\hat{G}^R_0(\pi)+\hat{G}^A_0(\pi)\right),
  \label{eq:G^<:E:I}\\
  \hat{G}^<_{\mib{E},II}(\pi)&=&\hat{G}^A_{\mib{E}}(\pi)-\hat{G}^R_{\mib{E}}(\pi),
  \label{eq:G^<:E:II}\\
  \hat{\Sigma}^<_{\mib{E},II}(\pi)&=&\hat{\Sigma}^A_{\mib{E}}(\pi)-\hat{\Sigma}^R_{\mib{E}}(\pi),
  \label{eq:Sigma^<:E,II}
\end{eqnarray}
and
\begin{eqnarray}
  \hat{G}^<_{\mib{B},I}(\pi) &=& 0,
  \label{eq:G^<:B:I}\\
  \hat{\Sigma}^<_{\mib{B},I}(\pi) &=& 0,
  \label{eq:Sigma^<:B:I}\\
  \hat{G}^<_{\mib{B},II}(\pi) &=& \hat{G}^A_{\mib{B}}(\pi)-\hat{G}^R_{\mib{B}}(\pi),
  \label{eq:G^<:B:II}\\
  \hat{\Sigma}^<_{\mib{B},II}(\pi) &=& \hat{\Sigma}^A_{\mib{B}}(\pi)-\hat{\Sigma}^R_{\mib{B}}(\pi).
  \label{eq:Sigma^<:B:II}
\end{eqnarray}

When $\hat{G}^<_{\mib{E},I}(\pi)$ is numerically calculated, Eq.~(\ref{eq:G^<:E:I}) should be used as it is. However, when it can be calculated analytically, the calculation carried out to solve Eq.~(\ref{eq:G^<:E:I}) is facilitated by constructing the following equation from Eq.~(\ref{eq:G^<:E:I}):
\begin{eqnarray}
  \lefteqn{\left[\hat{G}^<_{\mib{E},I}(\pi),\hat{H}_0(\mib{\pi})\right]_-+\hat{G}^<_{\mib{E},I}(\pi)\hat{\Sigma}^A_0(\pi)-\hat{\Sigma}^R_0(\pi)\hat{G}^<_{\mib{E},I}(\pi)}
  \nonumber\\
  &=&\hat{\Sigma}^<_{\mib{E},I}(\pi)\hat{G}^A_0(\pi)-\hat{G}^R_0(\pi)\hat{\Sigma}^<_{\mib{E},I}(\pi)
  \nonumber\\
  &&{}+\frac{i}{2}\left(
  (\hat{\Sigma}^A_0(\pi)-\hat{\Sigma}^R_0(\pi))(\mib{\nabla}_\pi\hat{G}^A_0(\pi))
  -(\mib{\nabla}_\pi(\hat{H}_0(\mib{\pi})+\hat{\Sigma}^R_0(\pi)))(\hat{G}^A_0(\pi)-\hat{G}^R_0(\pi))
  \right.\nonumber\\
  &&\left.\ \ \ \ {}+(\mib{\nabla}_\pi\hat{G}^R_0(\pi))(\hat{\Sigma}^A_0(\pi)-\hat{\Sigma}^R_0(\pi))-(\hat{G}^A_0(\pi)-\hat{G}^R_0(\pi))(\mib{\nabla}_\pi(\hat{H}_0(\mib{\pi})+\hat{\Sigma}^A_0(\pi)))\right).
  \nonumber\\
  \label{eq:G^<:E:I:QBE}
\end{eqnarray}
This is the so-called quantum Boltzmann equation for the classical
contribu-\linebreak tion~\cite{Mahan,RammerSmith86,HaugJauho} in the context of the Keldysh formalism or the Baym-Kadanoff equation~\cite{KadanoffBaym,HaugJauho} linearized in the electric field. This self-consistent equation~(\ref{eq:G^<:E:I:QBE}) can also be derived by substituting Eqs.~(\ref{eq:_G_:exp}) and (\ref{eq:_Sigma_:exp}) into Eq.~(\ref{eq:_QBE_:total}), linearizing it in $F^{\mu\nu}$, substituting Eqs.~(\ref{eq:G,Sigma^<:E:I,II}), and taking the classical terms proportional to $\partial_{\pi^0}f_\mp(\pi^0)$.

The classical contribution to the $\mib{E}$-linear lesser self-energy is given by
\begin{equation}
  \Sigma^<_{\mib{E},I}(\pi^0,\mib{\pi})=n_{\text{imp}}\int\!\frac{d\mib{\pi}'}{(2\pi\hbar)^d}\hat{u}_{\mib{\pi}-\mib{\pi}'}\hat{G}^<_{\mib{E},I}(\pi^0,\mib{\pi}')\hat{u}_{\mib{\pi}'-\mib{\pi}}
  \label{eq:Sigma^<:E:I:Born_1}
\end{equation}
in the Born approximation for the general impurity potential, Eq.~(\ref{eq:u_q}), and by
\begin{equation}
  \Sigma^<_{\mib{E},I}(\pi^0)=n_{\text{imp}}\hat{T}^R_0(\pi^0)\hat{g}^<_{\mib{E},I}(\pi^0)\hat{T}^A_0(\pi^0)
  \label{eq:Sigma^<:E:I:Botn_all}
\end{equation}
in the $T$-matrix approximation for the $\delta$-function potential, Eq.~(\ref{eq:u:delta}).

In conclusion, the problem at hand of obtaining the quantum distribution functions $\hat{G}^<_{\mib{E}}(\pi)$ and $\hat{G}^<_{\mib{B}}(\pi)$ is reduced to solving the self-consistent equation~(\ref{eq:G^<:E:I:QBE}) for the classical contribution $\hat{G}^<_{\mib{E},I}(\pi)$ and the self-consistent equations~(\ref{eq:G^R,A:E}) and (\ref{eq:G^R,A:B}) for the quantum contributions $\hat{G}^{R(A)}_{\mib{E}}(\pi)$ and $\hat{G}^{R(A)}_{\mib{B}}(\pi)$ over the occupied states, together with the diagrammatic representation of the self-energy according to the Feynmann rule~\cite{RammerSmith86}.

\section{Second-order response}
\label{sec:2nd}

To go beyond the linear response response to the external fields, we proceed to the second-order terms in Eq.~(\ref{eq:_G_:grad}).
In particular, substituting Eqs.~(\ref{eq:_G_:exp}) and (\ref{eq:_Sigma_:exp}) into Eq.~(\ref{eq:_G_:grad}) and comparing the second-order terms in $F$, we obtain
\begin{subequations}
\begin{eqnarray}
  \underline{\hat{G}}_{\mu\nu,\mu'\nu'}(\pi)
  &=&\underline{\hat{G}}_0(\pi)\left[
    \underline{\hat{\Sigma}}_{\mu\nu,\mu'\nu'}(\pi)\underline{\hat{G}}_0(\pi)
    +\underline{\hat{\Sigma}}_{\mu\nu}(\pi)\underline{\hat{G}}_{\mu'\nu'}(\pi)
    +\underline{\hat{\Sigma}}_{\mu'\nu'}(\pi)\underline{\hat{G}}_{\mu\nu}(\pi)
    \right.
    \nonumber\\
    &&\ \ \ \ \ {}+\frac{i}{2}\left\{\left(
    (\partial_{\pi^\mu}\underline{\hat{\Sigma}}_{\mu'\nu'}(\pi))(\partial_{\pi^\nu}\underline{\hat{G}}_0(\pi))
    -(\partial_{\pi^\mu}\underline{\hat{G}}_0^{-1}(\pi))(\partial_{\pi^\nu}\underline{\hat{G}}_{\mu'\nu'}(\pi))\right)\right.
    \nonumber\\
    &&\ \ \ \ \ \ \ \ \ \ {}-\left(\mu\leftrightarrow\nu\right)+((\mu\nu)\leftrightarrow(\mu'\nu'))-((\mu\nu\mu'\nu')\leftrightarrow(\nu'\mu'\mu\nu))\Bigr\}
    \nonumber\\
    &&\ \ \ \ \ {}+\frac{1}{4}\left((\partial_{\pi^\mu}\partial_{\pi^{\mu'}}\underline{\hat{G}}_0^{-1}(\pi))(\partial_{\pi^\nu}\partial_{\pi^{\nu'}}\underline{\hat{G}}_0(\pi))\right.
    \nonumber\\
    &&\ \ \ \ \ \ \ \ \ \ {}-(\mu\leftrightarrow\nu)-(\mu'\leftrightarrow\nu')+((\mu\mu')\leftrightarrow(\nu\nu'))\Bigr)\Bigr]
  \label{eq:_G_1:grad2}\\
  &=&\left[
    \underline{\hat{G}}_0(\pi)\underline{\hat{\Sigma}}_{\mu\nu,\mu'\nu'}(\pi)
    +\underline{\hat{G}}_{\mu\nu}(\pi)\underline{\hat{\Sigma}}_{\mu'\nu'}(\pi)
    +\underline{\hat{G}}_{\mu'\nu'}(\pi)\underline{\hat{\Sigma}}_{\mu\nu}(\pi)
    \right.
    \nonumber\\
    &&\ \ \ \ \ {}+\frac{i}{2}\left\{\left(
    (\partial_{\pi^\mu}\underline{\hat{G}}_0(\pi))(\partial_{\pi^\nu}\underline{\hat{\Sigma}}_{\mu'\nu'}(\pi))
    -(\partial_{\pi^\mu}\underline{\hat{G}}_{\mu'\nu'}(\pi))(\partial_{\pi^\nu}\underline{\hat{G}}_0^{-1}(\pi))\right)\right.
    \nonumber\\
    &&\ \ \ \ \ \ \ \ \ \ {}-\left(\mu\leftrightarrow\nu\right)+((\mu\nu)\leftrightarrow(\mu'\nu'))-((\mu\nu\mu'\nu')\leftrightarrow(\nu'\mu'\mu\nu))\Bigr\}
    \nonumber\\
    &&\ \ \ \ \ {}+\frac{1}{4}\left((\partial_{\pi^\mu}\partial_{\pi^{\mu'}}\underline{\hat{G}}_0(\pi))(\partial_{\pi^\nu}\partial_{\pi^{\nu'}}\underline{\hat{G}}_0^{-1}(\pi))\right.
    \nonumber\\
    &&\ \ \ \ \ \ \ \ \ \ {}-(\mu\leftrightarrow\nu)-(\mu'\leftrightarrow\nu')+((\mu\mu')\leftrightarrow(\nu\nu'))\Bigr)\Bigr]\underline{\hat{G}}_0(\pi).
  \label{eq:_G_2:grad2}
\end{eqnarray}
  \label{eq:_G_:grad2}%
\end{subequations}
In this section, the argument $\pi$ of the Green's functions and the self-energies is often suppressed.

\subsection{Magneto-transport properties}
\label{subsec:magneto-transport}

Now we consider the magneto-transport properties as a second-order response expressed by the cross terms of $\mib{E}$ and $\mib{B}$. We start from Eqs.~(\ref{eq:_G_:grad2}), which can be rewritten as
\begin{subequations}
\begin{eqnarray}
  \underline{\hat{G}}_{E_i,B_j}
  &=&\underline{\hat{G}}_0\biggl[
    \underline{\hat{\Sigma}}_{E_i,B_j}\underline{\hat{G}}_0
    +\underline{\hat{\Sigma}}_{E_i}\underline{\hat{G}}_{B_j}
    +\underline{\hat{\Sigma}}_{B_j}\underline{\hat{G}}_{E_i}
    \nonumber\\
    &&{}+\frac{i}{2}\Bigl\{\left(
    (\partial_{\pi^0}\underline{\hat{\Sigma}}_{B_j})(\partial_{\pi^i}\underline{\hat{G}}_0)
    -(\partial_{\pi^0}\underline{\hat{G}}_0^{-1})(\partial_{\pi^i}\underline{\hat{G}}_{B_j})\right)
    \nonumber\\
    &&\ \ \ \ \ {}-\left(
    (\partial_{\pi^i}\underline{\hat{\Sigma}}_{B_j})(\partial_{\pi^0}\underline{\hat{G}}_0)
    -(\partial_{\pi^i}\underline{\hat{G}}_0^{-1})(\partial_{\pi^0}\underline{\hat{G}}_{B_j})\right)
    \nonumber\\
    &&\ \ \ \ \ {}+\left(
    (\mib{\nabla}_\pi\underline{\hat{\Sigma}}_{E_i})\times(\mib{\nabla}_\pi\underline{\hat{G}}_0)
    -(\mib{\nabla}_\pi\underline{\hat{G}}_0^{-1})\times(\mib{\nabla}_\pi\underline{\hat{G}}_{E_i})\right)_j\Bigr\}
    \nonumber\\
    &&{}+\frac{1}{4}\left((\partial_{\pi^0}\mib{\nabla}_\pi\underline{\hat{G}}_0^{-1}))\times(\partial_{\pi^i}\mib{\nabla}_\pi\underline{\hat{G}}_0)
    -(\partial_{\pi^i}\mib{\nabla}_\pi(\underline{\hat{G}}_0^{-1}))\times(\partial_{\pi^0}\mib{\nabla}_\pi\underline{\hat{G}}_0)\right)_j\biggr]
  \nonumber\\
  \label{eq:_G_1:E,B}\\
  &=&\biggl[
    \underline{\hat{G}}_0\underline{\hat{\Sigma}}_{E_i,B_j}
    +\underline{\hat{G}}_{B_j}\underline{\hat{\Sigma}}_{E_i}
    +\underline{\hat{G}}_{E_i}\underline{\hat{\Sigma}}_{B_j}
    \nonumber\\
    &&{}+\frac{i}{2}\Bigl\{\left(
    (\partial_{\pi^0}\underline{\hat{G}}_0)(\partial_{\pi^i}\underline{\hat{\Sigma}}_{B_j})
    -(\partial_{\pi^0}\underline{\hat{G}}_{B_j})(\partial_{\pi^i}\underline{\hat{G}}_0^{-1})\right)
    \nonumber\\
    &&\ \ \ \ \ {}-\left(
    (\partial_{\pi^i}\underline{\hat{G}}_0)(\partial_{\pi^0}\underline{\hat{\Sigma}}_{B_j})
    -(\partial_{\pi^i}\underline{\hat{G}}_{B_j})(\partial_{\pi^0}\underline{\hat{G}}_0^{-1})\right)
    \nonumber\\
    &&\ \ \ \ \ {}+\left(
    (\mib{\nabla}_\pi\underline{\hat{G}}_0)\times(\mib{\nabla}_\pi\underline{\hat{\Sigma}}_{E_i})
    -(\mib{\nabla}_\pi\underline{\hat{G}}_{E_i})\times(\mib{\nabla}_\pi\underline{\hat{G}}_0^{-1})\right)_j\Bigr\}
    \nonumber\\
    &&\left.{}+\frac{1}{4}\left((\partial_{\pi^0}\mib{\nabla}_\pi\underline{\hat{G}}_0)\times(\partial_{\pi^i}\mib{\nabla}_\pi\underline{\hat{G}}_0^{-1})
    -(\partial_{\pi^i}\mib{\nabla}_\pi\underline{\hat{G}}_0)\times(\partial_{\pi^0}\mib{\nabla}_\pi\underline{\hat{G}}_0^{-1})\right)_j\right]\underline{\hat{G}}_0.
  \nonumber\\
  \label{eq:_G_2:E,B}
\end{eqnarray}
\label{eq:_G_:E,B}
\end{subequations}

The retarded (advanced) component can be written as
\begin{eqnarray}
  \hat{G}^{R(A)}_{E_i,B_j}
  &=&\hat{G}^{R(A)}_0\biggl[
    \hat{\Sigma}^{R(A)}_{E_i,B_j}\hat{G}^{R(A)}_0
    +\hat{\Sigma}^{R(A)}_{E_i}\hat{G}^{R(A)}_{B_j}
    +\hat{\Sigma}^{R(A)}_{B_j}\hat{G}^{R(A)}_{E_i}
    \nonumber\\
    &&{}+\frac{i}{2}\Bigl\{\left(
    (\partial_{\pi^0}\hat{\Sigma}^{R(A)}_{B_j})(\partial_{\pi^i}\hat{G}^{R(A)}_0)-(\partial_{\pi^0}\hat{G}^{R(A)-1}_0)(\partial_{\pi^i}\hat{G}^{R(A)}_{B_j})\right)
    \nonumber\\
    &&\ \ \ \ \ \ {}-\left(
    (\partial_{\pi^i}\hat{\Sigma}^{R(A)}_{B_j})(\partial_{\pi^0}\hat{G}^{R(A)}_0)-(\partial_{\pi^i}\hat{G}^{R(A)-1}_0)(\partial_{\pi^0}\hat{G}^{R(A)}_{B_j})\right)
    \nonumber\\
    &&\ \ \ \ \ \ {}+\left(
    (\mib{\nabla}_\pi\hat{\Sigma}^{R(A)}_{E_i})\times(\mib{\nabla}_\pi\hat{G}^{R(A)}_0)-(\mib{\nabla}_\pi\hat{G}^{R(A)-1}_0)\times(\mib{\nabla}_\pi\hat{G}^{R(A)}_{E_i})\right)_j\Bigr\}
    \nonumber\\
    &&{}+\frac{1}{4}\left((\partial_{\pi^0}\mib{\nabla}_\pi\hat{G}^{R(A)-1}_0)\times(\partial_{\pi^i}\mib{\nabla}_\pi\hat{G}^{R(A)}_0)\right.
    \nonumber\\
    &&\left.\left.\ \ \ \ \ \ \ {}-(\partial_{\pi^i}\mib{\nabla}_\pi(\hat{G}^{R(A)-1}_0))\times(\partial_{\pi^0}\mib{\nabla}_\pi\hat{G}^{R(A)}_0)\right)_j\right].
  \label{eq:G^R,A:E,B}
\end{eqnarray}

We know that $\hat{G}^<_{E_i}(\pi)$, $\hat{\Sigma}^<_{E_i}(\pi)$, $\hat{G}^<_{B_j}(\pi)$ and $\hat{\Sigma}^<_{B_j}(\pi)$ have the forms given in Eqs.~(\ref{eq:G^<:E:I,II}), (\ref{eq:Sigma^<:E:I,II}), (\ref{eq:G^<:B:I,II}) and (\ref{eq:Sigma^<:B:I,II}), supplemented by Eqs.~(\ref{eq:G^R,A:E}), (\ref{eq:G^R,A:B}), and (\ref{eq:G^<:E:I}). Therefore, using the property given by Eq.~(\ref{eq:app:U^<}), we find that the lesser component of Eq.~(\ref{eq:_G_:E,B}) also has the solution
\begin{subequations}
\begin{eqnarray}
  \hat{G}^<_{E_i,B_j}(\pi)&=&\hat{G}^<_{E_i,B_j,I}(\pi)\partial_{\pi^0}f_\mp(\pi^0)+\hat{G}^<_{E_i,B_j,II}(\pi)f_\mp(\pi^0),
  \label{eq:G^<:E,B:I,II}\\
  \hat{\Sigma}^<_{E_i,B_j}(\pi)&=&\hat{\Sigma}^<_{E_i,B_j,I}(\pi)\partial_{\pi^0}f_\mp(\pi^0)+\hat{\Sigma}^<_{E_i,B_j,II}(\pi)f_\mp(\pi^0),
  \label{eq:Sigma^<:E,B:I,II}
\end{eqnarray}
\label{eq:G,Sigma^<:E,B:I,II}%
\end{subequations}
with the self-consistent equation for $\hat{G}^<_{E_i,B_j,I}$ and $\Sigma^<_{E_i,B_j,I}$,
\begin{subequations}
\begin{eqnarray}
  \hat{G}^<_{E_i,B_j,I}
  &=&\hat{G}^R_0\biggl[\hat{\Sigma}^<_{E_i,B_j,I}\hat{G}^A_0+\hat{\Sigma}^<_{E_i,I}\hat{G}^A_{B_j}+\hat{\Sigma}^R_{B_j}\hat{G}^<_{E_i,I}
  \nonumber\\
  &&{}+\frac{i}{2}\Bigl\{
  (\hat{\Sigma}^A_{B_j}-\hat{\Sigma}^R_{B_j})(\partial_{\pi^i}\hat{G}^A_0)
  +(\hat{\Sigma}^A_0-\hat{\Sigma}^R_0)(\partial_{\pi^i}\hat{G}^A_{B_j})
  \nonumber\\
  &&\ \ \ \ \ {}-(\partial_{\pi^i}\hat{\Sigma}^R_{B_j})(\hat{G}^A_0-\hat{G}^R_0)
  -(\partial_{\pi^i}(\hat{H}_0(\mib{\pi})+\hat{\Sigma}^R_0))(\hat{G}^A_{B_j}-\hat{G}^R_{B_j})
  \nonumber\\
  &&\ \ \ \ \ {}+\left(
  (\mib{\nabla}_\pi\hat{\Sigma}^<_{E_i,I})\times(\mib{\nabla}_\pi\hat{G}^A_0)
  +(\mib{\nabla}_\pi(\hat{H}_0(\mib{\pi})+\hat{\Sigma}^R_0))\times(\mib{\nabla}_\pi\hat{G}^<_{E_i,I})\right)_j\Bigr\}
    \nonumber\\
    &&{}-\frac{1}{4}\left((\mib{\nabla}_\pi(\hat{\Sigma}^A_0-\hat{\Sigma}^R_0))\times(\partial_{\pi^i}\mib{\nabla}_\pi\hat{G}^A_0)\right.
    \nonumber\\
    &&\left.\left.\ \ \ \ {}-(\partial_{\pi^i}\mib{\nabla}_\pi(\hat{H}_0(\mib{\pi})+\hat{\Sigma}^R_0))\times(\mib{\nabla}_\pi(\hat{G}^A_0-\hat{G}^R_0))\right)_j\right]
  \label{eq:G^<1:E,B:I}\\
  &=&\biggl[\hat{G}^R_0\hat{\Sigma}^<_{E_i,B_j,I}
    +\hat{G}^R_{B_j}\hat{\Sigma}^<_{E_i,I}
    +\hat{G}^<_{E_i,I}\hat{\Sigma}^A_{B_j}
  \nonumber\\
  &&{}+\frac{i}{2}\left\{
  (\hat{G}^A_0-\hat{G}^R_0)(\partial_{\pi^i}\hat{\Sigma}^A_{B_j})
  +(\hat{G}^A_{B_j}-\hat{G}^R_{B_j})(\partial_{\pi^i}(\hat{H}_0(\mib{\pi})+\hat{\Sigma}^A_0))
  \right.\nonumber\\
  &&\ \ \ \ \ {}-(\partial_{\pi^i}\hat{G}^R_0)(\hat{\Sigma}^A_{B_j}-\hat{\Sigma}^R_{B_j})
  -(\partial_{\pi^i}\hat{G}^R_{B_j})(\hat{\Sigma}^A_0-\hat{\Sigma}^R_0)
  \nonumber\\
  &&\ \ \ \ \ {}+\left(
  (\mib{\nabla}_\pi\hat{G}^R_0)\times(\mib{\nabla}_\pi\hat{\Sigma}^<_{E_i,I})+(\mib{\nabla}_\pi\hat{G}^<_{E_i,I})\times(\mib{\nabla}_\pi(\hat{H}_0(\mib{\pi})+\hat{\Sigma}^A_0))\right)_j\Bigr\}
    \nonumber\\
    &&{}-\frac{1}{4}\left((\mib{\nabla}_\pi(\hat{G}^A_0-\hat{G}^R_0))\times(\partial_{\pi^i}\mib{\nabla}_\pi(\hat{H}_0(\mib{\pi})+\hat{\Sigma}^R_0))
    \right.\nonumber\\
    &&\left.\left.\ \ \ \ {}-(\partial_{\pi^i}\mib{\nabla}_\pi\hat{G}^R_0)\times(\mib{\nabla}_\pi(\hat{\Sigma}^A_0-\hat{\Sigma}^R_0))\right)_j\right]\hat{G}^A_0,
  \label{eq:G^<2:E,B:I}
\end{eqnarray}
  \label{eq:G^<:E,B:I}%
\end{subequations}
and
\begin{subequations}
\begin{eqnarray}
  \hat{G}^<_{E_i,B_j,II}(\pi)&=&\hat{G}^A_{E_i,B_j}(\pi)-\hat{G}^R_{E_i,B_j}(\pi),
  \label{eq:G^<:E,B,II}\\
  \hat{\Sigma}^<_{E_i,B_j,II}(\pi)&=&\hat{\Sigma}^A_{E_i,B_j}(\pi)-\hat{\Sigma}^R_{E_i,B_j}(\pi).
  \label{eq:Sigma^<:E,B,II}
\end{eqnarray}
\label{eq:G,Sigma^<:E,B,II}%
\end{subequations}

Equation~(\ref{eq:G^<:E,B:I}) can be transformed into an analogue of the quantum Boltzmann equation;
\begin{eqnarray}
  \lefteqn{\left[\hat{G}^<_{E_i,B_j,I},\hat{H}_0(\mib{\pi})\right]_-+\hat{G}^<_{E_i,B_j,I}\hat{\Sigma}^A_0-\hat{\Sigma}^R_0\hat{G}^<_{E_i,B_j,I}}
  \nonumber\\
  &=&\hat{\Sigma}^<_{E_i,B_j,I}\hat{G}^A_0-\hat{G}^R_0\hat{\Sigma}^<_{E_i,B_j,I}+\hat{\Sigma}^<_{E_i,I}\hat{G}^A_{B_j}-\hat{G}^R_{B_j}\hat{\Sigma}^<_{E_i,I}
  +\hat{\Sigma}^R_{B_j}\hat{G}^<_{E_i,I}-\hat{G}^<_{E_i,I}\hat{\Sigma}^A_{B_j}
  \nonumber\\
  &&{}+\frac{i}{2}\left\{
  (\hat{\Sigma}^A_{B_j}-\hat{\Sigma}^R_{B_j})(\partial_{\pi^i}\hat{G}^A_0)
  +(\partial_{\pi^i}\hat{G}^R_0)(\hat{\Sigma}^A_{B_j}-\hat{\Sigma}^R_{B_j})\right.
  \nonumber\\
  &&\ \ \ \ \ {}+(\hat{\Sigma}^A_0-\hat{\Sigma}^R_0)(\partial_{\pi^i}\hat{G}^A_{B_j})
  +(\partial_{\pi^i}\hat{G}^R_{B_j})(\hat{\Sigma}^A_0-\hat{\Sigma}^R_0)
  \nonumber\\
  &&\ \ \ \ \ {}-(\hat{G}^A_0-\hat{G}^R_0)(\partial_{\pi^i}\hat{\Sigma}^A_{B_j})
  -(\partial_{\pi^i}\hat{\Sigma}^R_{B_j})(\hat{G}^A_0-\hat{G}^R_0)
  \nonumber\\
  &&\ \ \ \ \ {}-(\hat{G}^A_{B_j}-\hat{G}^R_{B_j})(\partial_{\pi^i}(\hat{H}_0(\mib{\pi})+\hat{\Sigma}^A_0))
  -(\partial_{\pi^i}(\hat{H}_0(\mib{\pi})+\hat{\Sigma}^R_0))(\hat{G}^A_{B_j}-\hat{G}^R_{B_j})
  \nonumber\\
  &&\ \ \ \ \ {}+\left(
  (\mib{\nabla}_\pi\hat{\Sigma}^<_{E_i,I})\times(\mib{\nabla}_\pi\hat{G}^A_0)
  +(\mib{\nabla}_\pi(\hat{H}_0(\mib{\pi})+\hat{\Sigma}^R_0))\times(\mib{\nabla}_\pi\hat{G}^<_{E_i,I})\right)_j
    \nonumber\\
  &&\ \ \ \ \ {}-\left(
  (\mib{\nabla}_\pi\hat{G}^R_0)\times(\mib{\nabla}_\pi\hat{\Sigma}^<_{E_i,I})
  +(\mib{\nabla}_\pi\hat{G}^<_{E_i,I})\times(\mib{\nabla}_\pi(\hat{H}_0(\mib{\pi})+\hat{\Sigma}^A_0))\right)_j\Bigr\}
    \nonumber\\
    &&{}-\frac{1}{4}\left((\mib{\nabla}_\pi(\hat{\Sigma}^A_0-\hat{\Sigma}^R_0))\times(\partial_{\pi^i}\mib{\nabla}_\pi\hat{G}^A_0)
+(\partial_{\pi^i}\mib{\nabla}_\pi\hat{G}^R_0)\times(\mib{\nabla}_\pi(\hat{\Sigma}^A_0-\hat{\Sigma}^R_0))
    \right.\nonumber\\
    &&\ \ \ \ {}-(\mib{\nabla}_\pi(\hat{G}^A_0-\hat{G}^R_0))\times(\partial_{\pi^i}\mib{\nabla}_\pi(\hat{H}_0(\mib{\pi})+\hat{\Sigma}^A_0))
    \nonumber\\
    &&\left.\ \ \ \ {}-(\partial_{\pi^i}\mib{\nabla}_\pi(\hat{H}_0(\mib{\pi})+\hat{\Sigma}^R_0))\times(\mib{\nabla}_\pi(\hat{G}^A_0-\hat{G}^R_0))\right)_j.
      \label{eq:G^<:E,B:I:QBE}
\end{eqnarray}

\section{Transport properties in the single-band model}
\label{sec:transport_1}

In the single-band case, all the Green's functions $\hat{G}^{R,A,<}$ and the self-energies $\hat{\Sigma}^{R,A,<}$ are scalar quantities, and can be denoted as $G^{R,A,<}$ and $\Sigma^{R,A,<}$, respectively. Then, from Eqs.~(\ref{eq:G^R,A:E}) and (\ref{eq:G^R,A:B}), it is readily shown that $G^{R,A}_{\mib{E}}(\pi)=\Sigma^{R,A}_{\mib{E}}(\pi)=0$ and $G^{R,A}_{\mib{B}}(\pi)=\Sigma^{R,A}_{\mib{B}}(\pi)=0$, and hence the quantum contributions from $G^<_{\mib{E},II}$ to the electric current up to the linear response vanish. There exists only the classical contribution from $G^<_{\mib{E},I}$ in proportion to $\partial_{\pi^0} f_+(\pi^0)$, which becomes $\delta(\pi^0-\mu)$ at the zero temperature. Therefore, one has only to obtain $G^<_{\mib{E},I}(\mu,\mib{p}_F)$ and $\Sigma^<_{\mib{E},I}(\mu,\mib{p}_F)$. Equation~(\ref{eq:G^<:E:I}) or (\ref{eq:G^<:E:I:QBE}) is reduced to
\begin{equation}
  G^<_{\mib{E},I}(\pi)
  =\frac{\Im G^R_0(\pi)}{\Im \Sigma^R_0(\pi)}
  \left[\Sigma^<_{\mib{E},I}(\pi)-i\mib{\nabla}_\pi(H_0(\mib{\pi})+\Re\Sigma^R_0(\pi))\right]
  +i\Re\mib{\nabla}_\pi G^R_0(\pi).
  \label{eq:G^<:E:I:1}%
\end{equation}
This is the Bethe-Salpeter equation for the vertex correction $\Sigma^<_{\mib{E},I}(\pi)$.
Similarly, for the magneto-transport, from Eq.~(\ref{eq:G^<:E,B:I}) we obtain
\begin{eqnarray}
  \lefteqn{G^<_{E_i,B_j,I}(\pi) = \frac{1}{\Im\Sigma^R_0(\pi)}
  \left[\Im G^R_0(\pi)\Sigma^<_{E_i,B_j,I}(\pi)
    \right.}\nonumber\\
    &&-\frac{\epsilon_{\ell\ell'j}}{2}
    \Bigl((\partial_\ell\Sigma^<_{E_i,I}(\pi))(\partial_{\ell'}\Re G^R_0(\pi))
    +(\partial_\ell(H_0(\mib{\pi})+\Re\Sigma^R_0(\pi)))(\partial_{\ell'}G^<_{E_i,I}(\pi))
    \nonumber\\
  &&{}\left.-i\partial_i((\partial_\ell\Im\Sigma^R_0(\pi))(\partial_{\ell'}\Im G^R_0(\pi)))\Bigr)\right].
  \label{eq:G^<:E,B:I:1}
\end{eqnarray}
In the case of the magneto-transport, however, the quantum contributions characterized by $G^{R,A}_{E_i,B_j}(\pi)$ and $\Sigma^{R,A}_{E_i,B_j}(\pi)$ do not necessarily vanish unlike $G^{R,A}_{\mib{E}}(\pi)$, $\Sigma^{R,A}_{\mib{E}}(\pi)$, $G^{R,A}_{\mib{B}}(\pi)$, and $\Sigma^{R,A}_{\mib{B}}(\pi)$. From Eq.~(\ref{eq:G^R,A:E,B}), we can obtain the self-consistent equation to calculate $G^{R,A}_{E_i,B_j}(\pi)$ and $\Sigma^{R,A}_{E_i,B_j}(\pi)$,
\begin{eqnarray}
  G^{R(A)}_{E_i,B_j}(\pi)&=&G^{R(A)}_0(\pi)\Sigma^{R,A}_{E_i,B_j}(\pi)G^{R(A)}_0(\pi)
  \nonumber\\
  &&{}-\frac{1}{4}\left((\partial_{\pi^0}\mib{\nabla}_\pi \Sigma^{R(A)}_0(\pi))\times(\partial_{\pi^i}\mib{\nabla}_\pi G^{R(A)}_0(\pi))\right.
  \nonumber\\
  &&\left.\ \ \ \ {}-(\partial_{\pi^i}\mib{\nabla}_\pi(H_0(\mib{\pi})+\Sigma^{R(A)}_0(\pi)))\times(\partial_{\pi^0}\mib{\nabla}_\pi G^{R(A)}_0(\pi))\right)_j.
    \ \ \ 
    \label{eq:G^R,A:E,B:1}
\end{eqnarray}
In the following, we use the notation $(\varepsilon,\mib{p})$ instead of $(\pi^0,\mib{\pi})$.

Here, we employ the self-consistent Born approximation given by Eq.~(\ref{eq:_Sigma_:Born_1:exp}). More explicitly, we have
\begin{subequations}
\begin{eqnarray}
  \Sigma^R_0(\mu,\mib{p})&=&n_{\rm imp}\int\!\frac{d\mib{p}'}{(2\pi\hbar)^d}|u(\mib{p}-\mib{p}')|^2G^R_0(\mu,\mib{p}'),
  \label{eq:Sigma^R:0:1}\\
  \Sigma^<_{\mib{E},I}(\mu,\mib{p})&=&n_{\rm imp}\int\!\frac{d\mib{p}'}{(2\pi\hbar)^d}|u(\mib{p}-\mib{p}')|^2G^<_{\mib{E},I}(\mu,\mib{p}'),
  \label{eq:Sigma^<:E:I:1}\\
  \Sigma^<_{E_i,B_j,I}(\mu,\mib{p})&=&n_{\rm imp}\int\!\frac{d\mib{p}'}{(2\pi\hbar)^d}|u(\mib{p}-\mib{p}')|^2G^<_{E_i,B_j,I}(\mu,\mib{p}'),
  \label{eq:Sigma^<:E,B:I:1}\\
  \Sigma^{R,A}_{E_i,B_j}(\mu,\mib{p})&=&n_{\rm imp}\int\!\frac{d\mib{p}'}{(2\pi\hbar)^d}|u(\mib{p}-\mib{p}')|^2G^{R,A}_{E_i,B_j}(\mu,\mib{p}').
  \label{eq:Sigma^R,A:E,B:1}
\end{eqnarray}%
\end{subequations}

\subsection{Longitudinal conductivity}
\label{subsec:transport_1:E}

First, we calculate the longitudinal conductivity. Substituting Eq.~(\ref{eq:G^<:E:I:1}) into Eq.~(\ref{eq:Sigma^<:E:I:1}) and using Eq.~(\ref{eq:Sigma^R,A:E,B:1}), we obtain
\begin{eqnarray}
  &&\Sigma^<_{\mib{E},I}(\mu,\mib{p})-i\mib{\nabla}_p\Re\Sigma^R_0(\mu,\mib{p})
  \nonumber\\
  &=&\int\!\frac{d\mib{p}'}{(2\pi\hbar)^d}|u_{\mib{p}-\mib{p}'}|^2\frac{\Im G^R_0(\mu,\mib{p}')}{\Im\Sigma^R_0(\mu,\mib{p}')}
  \left[\Sigma^<_{\mib{E},I}(\mu,\mib{p}')
    -i\mib{\nabla}_{p'}(H_0(\mib{p}')+\Re\Sigma^R_0(\varepsilon,\mib{p}'))\right]
  \nonumber\\
  &=&-i(\tau_{\rm tr}/\tau-1)\mib{\nabla}_{p}H_0(\mib{p}),
  \label{eq:Sigma^<:E:I:sol}
\end{eqnarray}
and thus
\begin{equation}
  G^<_{\mib{E},I}(\mu,\mib{p})=-i\frac{\Im G^R_0(\mu,\mib{p})}{\Im \Sigma^R_0(\mu,\mib{p})}\frac{\tau_{\rm tr}}{\tau}\mib{\nabla}_{p}H_0(\mib{p})+i\mib{\nabla}_p \Re G^R_0(\mu,\mib{p}),
  \label{eq:G^<:E:I:sol}
\end{equation}
where we have assumed an isotropic dispersion $H_0(\mib{p})=\mib{p}^2/2m$ for simplicity, and introduced the electron lifetime $\tau$ and the transport lifetime $\tau_{\rm tr}$ via the relations
\begin{eqnarray}
  \hbar/\tau_{\rm tr}&=&2\int\!\frac{d\mib{p}'_F}{(2\pi\hbar)^d}|u_{\mib{p}_F-\mib{p}'_F}|^2(1-\cos\theta)(-\Im G^R_0(\mu,\mib{p}'_F))=\int\!d\Omega f(\Omega)(1-\cos\theta),
  \nonumber\\
  \label{eq:tau_tr}\\
  \hbar/\tau&=&2\int\!\frac{d\mib{p}'_F}{(2\pi\hbar)^d}|u_{\mib{p}_F-\mib{p}'_F}|^2(-\Im G^R_0(\mu,\mib{p}'_F))=\int\!d\Omega f(\Omega),
\end{eqnarray}
and the differential cross-section
\begin{equation}
  f(\Omega)=2\pi n_{\rm imp}\frac{m^* p_F^{d-2}\Omega_d}{(2\pi\hbar)^d}|u(\mib{p}_F-\mib{p}_F')|^2
  \label{eq:Born1:1:f}
\end{equation}
with the polar coordinates $\Omega=(\theta,\varphi)$ of $\mib{p}_F$ with respect to $\mib{p}'_F$ and $\Omega_d=\Gamma(d/2)/\pi^{d/2}$.
Here, $m^*$ is the effective mass given by
\begin{equation}
  1/m^*\equiv z/mb\equiv z(1/m+\nabla_p^2\Re\Sigma^R_0(\mu,p))|_{p\to p_F}
  \label{eq:m^*}
\end{equation}
with the renormalization factor defined by
\begin{equation}
  1/z\equiv 1-\partial_\varepsilon\Re\Sigma^R_0(\varepsilon,p_F)|_{\varepsilon\to\mu}.
  \label{eq:z}
\end{equation}

From Eqs.~(\ref{eq:G^<:E:I:1}) and (\ref{eq:Sigma^<:E:I:sol}), the conventional expression for the longitudinal electric conductivity at the zero temperature is reproduced as
\begin{eqnarray}
  \sigma_{ij}&=&\frac{e^2\hbar}{2\pi i}\int\!d\varepsilon\int\!\frac{d\mib{p}}{(2\pi\hbar)^d}(\partial_{p^i}H_0(p))G^<_{E_j}(\mu,\mib{p})\partial_\varepsilon f_+(\varepsilon)
  \nonumber\\
  &=&\frac{e^2\hbar}{2\pi}\int\!\frac{d\mib{p}}{(2\pi\hbar)^d}(\partial_{p^i}H_0(p)) \left[ G^R_0(\mu,\mib{p})(\partial_{p^j}H_0(p))G^A_0(\mu,\mib{p})\frac{\tau_{\text{tr}}}{\tau}
    -\partial_{p^j}\Re G^R_0(\mu,\mib{p})\right]
  \nonumber\\
  &=&\delta_{ij}\frac{n^*e^2\tau_{\text{tr}}}{m}\frac{m^*}{m}=\delta_{ij}\frac{n^*e^2\tau_{\text{tr}}^*}{m}b
  \label{eq:sigma_ij}
\end{eqnarray}
in the absence of the magnetic field. We have introduced $n^*=zn$ with the electron density
\begin{equation}
  n = \Omega_d p_F^d/d(2\pi\hbar)^d
  \label{eq:n}
\end{equation}
and the quasiparticle transport lifetime 
\begin{equation}
  \tau^*_{\text{tr}}=\tau_{\text{tr}}/z.
  \label{eq:tau_tr^*}
\end{equation}
It is important that the renormalization of the velocity due to the self-energy disappears in the above expression for $\sigma_{ii}$, in agreement with the Kubo formula. 

\subsection{Hall effect}
\label{subsec:transport_1:Hall}

We proceed to the second-order response of the electric current to the cross term of the electric field and the magnetic field. In the present model, using Eqs.~(\ref{eq:Sigma^<:E:I:sol}) and (\ref{eq:G^<:E:I:sol}), Eq.~(\ref{eq:G^<:E,B:I:1}) is reduced to
\begin{eqnarray}
  \lefteqn{G^<_{E_i,B_j,I}(\varepsilon,\mib{p})-i\epsilon_{ij\ell}\frac{\tau_{\text{tr}}}{\hbar m}(\partial_{p^\ell}\Re G^R_0(\varepsilon,\mib{p}))}
  \nonumber\\
  &=&G^R_0(\varepsilon,\mib{p})G^A_0(\varepsilon,\mib{p})\left[\Sigma^<_{E_i,B_j,I}(\varepsilon,\mib{p})-i\epsilon_{ij\ell}\frac{\tau_{\text{tr}}}{\hbar m}\partial_{p^\ell}(H_0(\mib{p})+\Re\Sigma^R_0(\varepsilon,\mib{p}))\right],
  \ \ \ 
  \label{eq:G^<:E,B:I:1_m}
\end{eqnarray}
as shown in Appendix~\ref{app:Derive}. Therefore, from Eqs.~(\ref{eq:Sigma^R:0:1}) and (\ref{eq:Sigma^<:E,B:I:1}), we obtain
\begin{eqnarray}
  \lefteqn{\Sigma^<_{E_i,B_j,I}(\mu,\mib{p})-i\epsilon_{ij\ell}\frac{\tau_{\text{tr}}}{\hbar m}\partial_{p^\ell}\Re\Sigma^R_0(\mu,\mib{p})}
  \nonumber\\
  &=&\int\!\frac{d\mib{p}'}{(2\pi\hbar)^d}|u_{\mib{p}-\mib{p}'}|^2\frac{\Im G^R_0(\mu,\mib{p}')}{\Im\Sigma^R_0(\mu,\mib{p}')}
  \nonumber\\
  &&\times
  \left[\Sigma^<_{E_i,B_j,I}(\mu,\mib{p}')
    -i\epsilon_{ij\ell}\frac{\tau_{\text{tr}}}{\hbar m}\partial_{{p'}^\ell}(H_0(\mib{p}')+\Re\Sigma^R_0(\mu,\mib{p}'))\right]
  \nonumber\\
  &=&-i\epsilon_{ij\ell}(\tau_{\rm tr}/\tau-1)\frac{\tau_{\text{tr}}}{\hbar m}(\partial_{p^\ell}H_0(p)),
  \label{eq:Sigma^<:E,B:I:sol}
\end{eqnarray}
and thus
\begin{eqnarray}
  G^<_{E_i,B_j,I}(\mu,\mib{p})&=&-i\epsilon_{ij\ell}\frac{\tau_{\text{tr}}}{\hbar m}\left[\frac{\Im G^R_0(\varepsilon,\mib{p})}{\Im\Sigma^R_0(\varepsilon,\mib{p})}\frac{\tau_{\text{tr}}}{\tau}(\partial_{p^\ell}H_0(p))
  -(\partial_{p^\ell}\Re G^R_0(\mu,\mib{p}))\right],
  \ \ \ \ \ 
  \label{eq:G^<:E,B:I:sol}
\end{eqnarray}
as also shown in Appendix~\ref{app:Derive}.
Consequently, we obtain the Fermi-surface contribution to the Hall
conductivity as
\begin{eqnarray}
  \sigma_{ij}^I&=&B_\ell\frac{e^3\hbar^2}{2\pi i}\int\!d\varepsilon\int\frac{d\mib{p}}{(2\pi\hbar)^d}(\partial_{p^i}H_0(\mib{p})) G^<_{E_j,B_\ell,I}(\varepsilon,\mib{p})\partial_\varepsilon f_+(\varepsilon)
  \nonumber\\
  &=&-B_\ell\frac{e^3\hbar^2}{2\pi i}\int\frac{d\mib{p}}{(2\pi\hbar)^d}(\partial_{p^i}H_0(\mib{p})) G^<_{E_j,B_\ell,I}(\mu,\mib{p})
  \nonumber\\
  &=&B^\ell\epsilon_{ij\ell}\frac{e^3\hbar^2}{2\pi}\int\frac{d\mib{p}}{(2\pi\hbar)^d}\frac{\tau_{\text{tr}}}{\hbar m}
  \nonumber\\
  &&{}\times(\partial_{p^i}H_0(\mib{p}))
  \left[G^R_0(\mu,\mib{p})G^A_0(\mu,\mib{p})\frac{\tau_{\text{tr}}}{\tau}(\partial_{p^i}H_0(p))
  -(\partial_{p^i}\Re G^R_0(\mu,\mib{p}))\right]
  \nonumber\\
  &=&\epsilon_{ij\ell}B^\ell\frac{n^*e^3\tau_{\rm tr}^2}{m^2}\frac{m^*}{m}=\epsilon_{ij\ell}B^\ell\sigma_{ii}^2\frac{1}{n^*e}\frac{m}{m^*}=\epsilon_{ij\ell}B^\ell\sigma_{ii}^2\frac{1}{neb}.
  \label{eq:sigma_H:I}
\end{eqnarray}

Next, we proceed to the quantum contribution determined by $G^{R,A}_{E_i,B_j}(\varepsilon,\mib{p})$. From Eq.~(\ref{eq:G^R,A:E,B:1}), we obtain
\begin{eqnarray}
  G^{R(A)}_{E_i,B_j}(\varepsilon,\mib{p})&=&{G^{R(A)}_0}^2(\varepsilon,\mib{p})\Sigma^{R,A}_{E_i,B_j}(\varepsilon,\mib{p})
  \nonumber\\
  &&{}-\frac{\epsilon_{ij\ell}}{4}\left((\partial_{p^i}^2(H_0(\mib{p})+\Sigma^{R(A)}_0(\varepsilon,\mib{p})))(\partial_\varepsilon\partial_{p^\ell} G^{R(A)}_0(\varepsilon,\mib{p}))
  \right.\nonumber\\
  &&\left.\ \ \ \ \ \ \ {}+(\partial_\varepsilon\partial_{p^\ell} \Sigma^{R(A)}_0(\pi))(\partial_{p^i}^2 G^{R(A)}_0(\pi))\right).
    \label{eq:G^R,A:E,B:1:sol}
\end{eqnarray}
In particular, in the case of a $\delta$-functional potential, $\partial_{p^i}\Sigma^R_0$ vanishes, and so it is reduced to
\begin{equation}
  G^{R(A)}_{E_i,B_j}(\varepsilon,\mib{p})=-\frac{\epsilon_{ij\ell}}{4m}
  G^{R(A)}_0(\varepsilon,\mib{p})\partial_\varepsilon\partial_{p^\ell}G^{R(A)}_0(\varepsilon,\mib{p}).
\end{equation}
Thus, we obtain the quantum contribution to the Hall conductivity as
\begin{eqnarray}
  \sigma_{ij}^{II}&=&B_\ell\frac{e^3\hbar^2}{2\pi i}\int\!d\varepsilon\int\frac{d\mib{p}}{(2\pi\hbar)^d}(\partial_{p^i}H_0(\mib{p})) (-2i\Im G^R_{E_j,B_\ell}(\varepsilon,\mib{p})f_+(\varepsilon))
  \nonumber\\
  &=&-B_\ell\frac{e^3\hbar^2}{2\pi}\int\!d\varepsilon\int\frac{d\mib{p}}{(2\pi\hbar)^d}\frac{1}{2m}(\partial_{p^i}^2H_0(\mib{p})) \Im G^R_0(\varepsilon,\mib{p})\partial_\varepsilon f_+(\varepsilon)
  \nonumber\\
  &=&B^\ell\epsilon_{ij\ell}\frac{e^3\hbar^2}{2\pi}\frac{1}{2m^2}\int\frac{d\mib{p}}{(2\pi\hbar)^d}\Im G^R_0(\mu,\mib{p}).
  \label{eq:sigma_H:II}
\end{eqnarray}
This term is negiligible in the usual metal compared with the classical contribution given by Eq.~(\ref{eq:sigma_H:I}).
The non-local self-energy produces a small correction to ths expression, which is of higher order in $1/\tau$. Therefore, it is negligible in the weak scattering limit.

From the above results, we reproduce the conventional expression for the Hall coefficient
\begin{equation}
  R=\rho_{yx}/B^z=\frac{1}{n^*e}\frac{m}{m^*}=\frac{1}{neb}.
  \label{eq:Hall}
\end{equation}

The above calculation gives an expansion of the Hall conductivity/resistivity in $\omega_c\tau_\text{tr}\ll1$ with the cyclotron frequency $\omega_c=eB/m$ and a weak magnetic field $B$. We also note that the above results can be easily extended to the case of a finite temperature $T$ by exploiting the Sommerfeld expansion and expanding the self-energies and the vertex corrections in $T$ to obtain the coefficients to the $T^2$ terms in the conductivity tensor and the Hall coefficient. 

\section{Discussion and conclusions}
\label{sec:discuss}

We have developed a new theoretical framework to study the electromagnetic response of the charged system in terms of the Keldysh formalism. The Moyal product and its expansion are the key concepts in this approach, and the gauge invariance is implemented from the beginning of the calculation. Here we discuss the relationship of the present approach to others, which are equivalent, in principle, within their domains of validity, but differ with regard to procedures of their calculations.

The most well-established formalism is the Kubo formula for the linear response, where the retarded response functions are obtained from the Matsubara Green's function via the analytic continuation, which can be calculated using the diagram method. In this formalism, it is necessary to take the diagrams for the self-energy and vertex correction consistently in order to satisfy the Ward-Takahashi identity. This is rather straightforward for the conductivity, but it requires an ingenious treatment for the Hall conductivity~\cite{Fukuyama69_1}. In the latter case, to guarantee the gauge invariance, the factor $q^2 \delta_{a b} - q_a q_b$ where $q_a$ is the wavenumber of the vector potential of the external magnetic field must be extracted and a careful treatment of the self-energy and vertex corrections is required, as mentioned above. In our formalism, by contrast, the equation of motion is gauge-invariant from the beginning, because of the replacement of the argument, $p \to \pi = p - q A$, and the fact that the vertex correction is represented by $G^<$ and $\Sigma^<$, while the self-energy by $\Sigma^{R,A}$. It is noted here that both $\Sigma^<$ and $\Sigma^{R,A}$ satisfy a single equation represented in the Keldysh space. Another advantage of the present approach is that no analytic continuation is needed. This is particularly useful when one analyzes non-linear responses, which is beyond the scope of the Kubo formalism.

  Another standard treatment of the non-equilibrium state is given by the Boltzmann transport theory. This is closely related to the Keldysh formalism in spirit, and $G^<$ is the quantum analogue of the distribution function $f(\mib{k},\mib{r})$ appearing in the semi-classical Boltzmann equation. Actually, the quantum Boltzmann equation is derived from the Keldysh formalism, and by assuming $G^<_0(\pi) = 2\pi i\delta(\pi^0- \varepsilon( \mib{\pi}) )f_{\mp}(\pi^0)$ with $\varepsilon(\mib{\pi})$ being the energy dispersion of the particle, it reduces to the semi-classical Boltzmann equation. This formalism is akin to the present one, but the change of the self-energy $\hat{\Sigma}^{R,A}$ and the Green's function $\hat{G}^{R,A}$ due to the external electromagnetic field is often ignored in the linear response regime. This semi-classical assumption is valid for single-band models, but not for multi-band systems, since in the latter case, the quantum contribution is missing in this assumption. Our formalism can also treat this multi-band case systematically.

  Lastly, it is interesting to note the relation of the present formalism to the St\u{r}eda formula~\cite{Streda82} for the linear response. St\u{r}eda decomposed the electric conductivity tensor $\sigma_{ij}$ into $\sigma_{ij}^I + \sigma_{ij}^{II}$, where the classical contribution $\sigma_{ij}^I$ comes from fermions at the Fermi surface while the quantum contribution $\sigma_{ij}^{II}$ from all the occupied states. Without the magnetic field, Eq.~(\ref{eq:G,Sigma^<:E:I,II}) with Eqs.~(\ref{eq:G^R,A:E}), (\ref{eq:G^<:E:I}), and (\ref{eq:G^<:E:II}) coincides with the St\u{r}eda formula~\cite{Streda82} in the absence of $\underline{\hat{\Sigma}}$. Even under a magnetic field, a similar decomposition of Eq.~(\ref{eq:G,Sigma^<:E,B:I,II}) facilitates the calculation of the conductivity, as explicitly shown in \S\ref{subsec:transport_1:Hall}. This decomposition is particularly useful when one investigates the intrinsic and extrinsic contributions to the Hall conductivity, an interpretation of which has been controversial for anomalous Hall effect for many decades but clarified in a unified way~\cite{Onoda_ahe}. Our present formalism provides a method to calculate $\sigma_{ij}^I$ and $\sigma_{ij}^{II}$ diagrammatically even in the presence of the particle-particle interaction. Therefore, the present approach can be regarded as a further development of the St\u{r}eda formula.

\section*{Acknowledgements}
The authors are grateful to H. Fukuyama and H. Kohno for valuable discussions.

\appendix
\section{Moyal Product}
\label{app:Moyal}

We start from the Wigner  mixed representation of $\underline{\hat{\Upsilon}}$ and $\underline{\hat{\Xi}}$ in terms of $(X;p)$:
\begin{subequations}
\begin{eqnarray}
  \underline{\hat{\Upsilon}}(X;p)&=&\int\!dx\,e^{ip_\mu x^\mu/\hbar}\underline{\hat{\Upsilon}}\left(\frac{X+x}{2},\frac{X-x}{2}\right),
  \label{eq:Upsilon:X;p}\\
  \underline{\hat{\Xi}}(X;p)&=&\int\!dx\,e^{ip_\mu x^\mu/\hbar}\underline{\hat{\Upsilon}}\left(\frac{X+x}{2},\frac{X-x}{2}\right).
  \label{eq:Xi:X;p}
\end{eqnarray}%
\end{subequations}
Then, the Fourier transform of the convolution given by Eq.~(\ref{eq:convolution:x_1;x_2}) reads
\begin{eqnarray}
  (\underline{\hat{\Upsilon}}*\underline{\hat{\Xi}})(X;p)
  &=&\int\!dx\,e^{-ip_\mu x^\mu/\hbar}\int\!dx_3\underline{\hat{\Upsilon}}\left(\frac{X+x}{2},\frac{X}{2}+x_3\right)\underline{\hat{\Xi}}\left(\frac{X}{2}+x_3,\frac{X-x}{2}\right)
  \nonumber\\
  &=&\int\!dx\int\!dx_3\int\!\frac{dp'}{(2\pi\hbar)^{d+1}}\,e^{ip'_\mu (\frac{x^\mu}{2}-x_3^\mu)/\hbar}\int\!\frac{dp''}{(2\pi\hbar)^{d+1}}\,e^{-ip''_\mu (x_3^\mu+\frac{x^\mu}{2})/\hbar}
  \nonumber\\
  &&\times
  \underline{\hat{\Upsilon}}\left(X+\frac{x}{4}+\frac{x_3}{2};p+p'\right)
  \underline{\hat{\Xi}}\left(X-\frac{x}{4}+\frac{x_3}{2};p-p''\right)
  \nonumber\\
  &=&\int\!dX'\int\!dX''\int\!\frac{dp'}{(2\pi\hbar)^{d+1}}\int\!\frac{dp''}{(2\pi\hbar)^{d+1}}\,e^{i(p'_\mu {X''}^\mu+p''_\mu {X'}^\mu)/\hbar}
  \nonumber\\
  &&\times 
  \underline{\hat{\Upsilon}}\left(X+\frac{X'}{2};p+p'\right)
  \underline{\hat{\Xi}}\left(X-\frac{X''}{2};p-p''\right)
  \nonumber\\
  &=&\int\!dX'\int\!dX''\int\!\frac{dp'}{(2\pi\hbar)^{d+1}}\int\!\frac{dp''}{(2\pi\hbar)^{d+1}}\,e^{i(p'_\mu {X''}^\mu+p''_\mu {X'}^\mu)/\hbar}
  \nonumber\\
  &&\times 
  \underline{\hat{\Upsilon}}(X;p)
  e^{\frac{{X'}^\mu}{2}\overleftarrow{\partial}_{X^\mu}+p'_\mu\overleftarrow{\partial}_{p_\mu}-\frac{{X''}^\mu}{2}\overrightarrow{\partial}_{X^\mu}-{p''}_\mu\overrightarrow{\partial}_{p_\mu}}
  \underline{\hat{\Xi}}(X;p)
  \nonumber\\
  &=&\int\!dX'\int\!dX''\int\!\frac{dp'}{(2\pi\hbar)^{d+1}}\int\!\frac{dp''}{(2\pi\hbar)^{d+1}}
  \nonumber\\
  &&\times \underline{\hat{\Upsilon}}(X;p)
  e^{i(p'_\mu/\hbar+\frac{i}{2}\overrightarrow{\partial}_{X^\mu})({X''}^\mu-i\hbar\overleftarrow{\partial}_{p_\mu})+i(p''_\mu/\hbar-\frac{i}{2}\overleftarrow{\partial}_{X^\mu})({X'}^\mu+i\hbar\overrightarrow{\partial}_{p_\mu})}
  \nonumber\\
  &&\times e^{\frac{i\hbar}{2}(\overleftarrow{\partial}_{X^\mu}\overrightarrow{\partial}_{p_\mu}-\overleftarrow{\partial}_{p_\mu}\overrightarrow{\partial}_{X^\mu})}
  \underline{\hat{\Xi}}(X;p)
  \nonumber\\
  &=&\underline{\hat{\Upsilon}}(X;p)
  e^{\frac{i\hbar}{2}(\overleftarrow{\partial}_{X^\mu}\overrightarrow{\partial}_{p_\mu}-\overleftarrow{\partial}_{p_\mu}\overrightarrow{\partial}_{X^\mu})}
  \underline{\hat{\Xi}}(X;p).
  \label{eq:app:Moyal}
\end{eqnarray}

\section{Basic Properties of the Retarded, Advanced, and Lesser Components of Product of the Matrix-Form Functions in the Keldysh Space}
\label{app:R,A,<}

Given the matrix-form functions in the Keldysh space,
\begin{equation}
  \underline{\hat{\Upsilon}}_i(p)=\left(\begin{array}{cc}
    \hat{\Upsilon}^R_i(p) & 2\hat{\Upsilon}^<_i(p)\\
    0                     &  \hat{\Upsilon}^A_i(p)
    \end{array}\right)
  \label{eq:app:U_i}
\end{equation}
with $i=1,\cdots,n$, their product 
\begin{equation}
  \underline{\hat{\Upsilon}}(p)=\underline{\hat{\Upsilon}}_1(p)\underline{\hat{\Upsilon}}_2(p)\cdots\underline{\hat{\Upsilon}}_n(p)
  \label{eq:app:U}
\end{equation}
satisfies the following relations:
\begin{eqnarray}
  \hat{\Upsilon}^R(p)&=&\hat{\Upsilon}^R_1(p)\hat{\Upsilon}^R_2(p)\cdots\hat{\Upsilon}^R_n(p),
  \label{eq:app:U^R}\\
  \hat{\Upsilon}^A(p)&=&\hat{\Upsilon}^A_1(p)\hat{\Upsilon}^A_2(p)\cdots\hat{\Upsilon}^A_n(p).
  \label{eq:app:U^A}
\end{eqnarray}
Furthermore, when each $\underline{\hat{\Upsilon}}_i$ satisfies the {\it equilibirum} condition
\begin{equation}
  \hat{\Upsilon}^<_i(p)=\left(\hat{\Upsilon}^A_i(p)-\hat{\Upsilon}^R_i(p)\right)f_\mp(\varepsilon),
  \label{eq:app:U^<_i}
\end{equation}
then the lesser component of the product is reduced to
\begin{eqnarray}
  \hat{\Upsilon}^<(p)
  &=&\hat{\Upsilon}^<_1\hat{\Upsilon}^A_2\cdots\hat{\Upsilon}^A_n
  \nonumber\\
  &&+\hat{\Upsilon}^R_1\hat{\Upsilon}^<_2\hat{\Upsilon}^A_3\cdots\hat{\Upsilon}^A_n
  \nonumber\\
  &&+\hat{\Upsilon}^R_1\hat{\Upsilon}^R_2\hat{\Upsilon}^<_3\hat{\Upsilon}^A_4\cdots\hat{\Upsilon}^A_n
  \nonumber\\
  &&+\cdots
  \nonumber\\
  &&+\hat{\Upsilon}^R_1\cdots\hat{\Upsilon}^R_{n-2}\hat{\Upsilon}^<_{n-1}\hat{\Upsilon}^A_n
  \nonumber\\
  &&+\hat{\Upsilon}^R_1\cdots\hat{\Upsilon}^R_{n-1}\hat{\Upsilon}^<_n
  \nonumber\\
  &=&\left(\hat{\Upsilon}^A(p)-\hat{\Upsilon}^R(p)\right)f_\mp(\varepsilon)
  \label{eq:app:U^<}
\end{eqnarray}
for $n\ge2$.

\section{Derivation of Eq.~(\ref{eq:G^<:E,B:I:sol})}
\label{app:Derive}

We start from Eq.~(\ref{eq:G^<:E,B:I:1}), which has been reduced from Eq.~(\ref{eq:G^<:E,B:I:QBE}) in the single-component scalar problem. Then, substituting Eqs.~(\ref{eq:G^<:E:I:sol}) and (\ref{eq:Sigma^<:E:I:sol}) into Eq.~(\ref{eq:G^<:E,B:I:1}) leads to
\begin{eqnarray}
  \lefteqn{G^<_{E_i,B_j,I}(\varepsilon,\mib{p})
  =\frac{1}{\Im\Sigma^R_0(\varepsilon,\mib{p})}
  \left[\Im G^R_0(\varepsilon,\mib{p})\Sigma^<_{E_i,B_j,I}(\varepsilon,\mib{p})
    \right.}\nonumber\\
    &&{}+i\frac{\epsilon_{\ell\ell'j}}{2}
    \left\{(\partial_{p^\ell}\frac{\tau_{\text{tr}}}{\tau}\partial_{p^i}H_0(\mib{p}))(\partial_{p^{\ell'}}\Re G^R_0(\varepsilon,\mib{p}))
    \right.\nonumber\\
    &&\left.\ \ \ \ \ \ \ \ \ {}+(\partial_{p^\ell}(H_0(\mib{p})+\Re\Sigma^R_0(\varepsilon,\mib{p})))(\partial_{p^{\ell'}}\frac{\tau_{\text{tr}}}{\tau}G^R_0(\varepsilon,\mib{p})G^A_0(\varepsilon,\mib{p})\partial_{p^i}H_0(\mib{p}))
    \right\}
    \nonumber\\
    &&{}-i\frac{\epsilon_{\ell\ell'j}}{2}
    \left\{(\partial_{p^\ell}\partial_{p^i}(H_0(\mib{p})+\Re\Sigma^R_0(\varepsilon,\mib{p})))(\partial_{p^{\ell'}}\Re G^R_0(\varepsilon,\mib{p}))
    \right.\nonumber\\
    &&\ \ \ \ \ \ \ \ \ {}+(\partial_{p^\ell}(H_0(\mib{p})+\Re\Sigma^R_0(\varepsilon,\mib{p})))(\partial_{p^{\ell'}}\partial_{p^i}\Re G^R_0(\varepsilon,\mib{p}))
    \nonumber\\
    &&\left.\left.\ \ \ \ \ \ \ \ \ {}-\partial_{p^i}\left((\partial_{p^\ell}\Im\Sigma^R_0(\mu,\mib{p}))(\partial_{p^{\ell'}}\Im G^R_0(\mu,\mib{p}))\right)\right\}    \right]
  \nonumber\\
  &=&\frac{1}{\Im\Sigma^R_0(\varepsilon,\mib{p})}
  \left[\Im G^R_0(\varepsilon,\mib{p})\Sigma^<_{E_i,B_j,I}(\varepsilon,\mib{p})
    \right.\nonumber\\
    &&+i\frac{\epsilon_{\ell\ell'j}}{2}\left\{
    (\partial_{p^\ell}\frac{\tau_{\text{tr}}}{\tau}\partial_{p^i}H_0(\mib{p}))
    (\partial_{p^{\ell'}}\Re G^R_0(\varepsilon,\mib{p}))
    \right.\nonumber\\
    &&\left.\ \ \ \ \ \ \ 
	  {}-(\partial_{p^\ell}\frac{\tau_{\text{tr}}}{\tau}\partial_{p^i}H_0(\mib{p}))
    G^R_0(\varepsilon,\mib{p})G^A_0(\varepsilon,\mib{p})(\partial_{\pi^{\ell'}}(H_0(\mib{p})+\Re\Sigma^R_0(\varepsilon,\mib{p})))\right)
    \nonumber\\
    &&\ \ \ \ \ \ \ {}-i\frac{\tau_{\text{tr}}}{\tau}(\Im\Sigma^R_0(\varepsilon,\mib{p}))(\partial_{p^i}H_0(\mib{p}))(\partial_{p^\ell}G^R_0(\varepsilon,\mib{p}))(\partial_{p^{\ell'}}G^A_0(\varepsilon,\mib{p}))
    \nonumber\\
    &&\left.\left.\ \ \ \ \ \ \ {}-\partial_{p^i}\Re\left((\partial_{p^\ell}(H_0(\mib{p})+\Sigma^R_0(\mu,\mib{p})))(\partial_{p^{\ell'}}G^R_0(\mu,\mib{p}))\right)\right\}\right].
  \label{eq:app:G^<:E,B:I:1_m}
\end{eqnarray}
The last line of the above equation identically vanishes due to the anti-symmetric tensor $\epsilon_{\ell\ell'j}$. Using $\epsilon_{\ell\ell'j}(\partial_{p^\ell}g_1(p))(\partial_{p^{\ell'}}g_2(p))=0$, Eq.~(\ref{eq:app:G^<:E,B:I:1_m}) can then be rewritten as Eq.~(\ref{eq:G^<:E,B:I:1_m}) in the present model $H_0(\mib{p})=\mib{p}^2/2m$.

Multiplying both sides of Eq.~(\ref{eq:G^<:E,B:I:1_m}) by $|u_{\mib{p}-\mib{p}'}|^2\hbar/\tau_{\text{tr}}$ and performing a momentum integration, we obtain
\begin{eqnarray}
  \lefteqn{\int\frac{d\mib{p}'}{(2\pi\hbar)^d}|u_{\mib{p}-\mib{p}'}|^2\left[\frac{\hbar}{\tau_{\text{tr}}}G^<_{E_i,B_j,I}(\varepsilon,\mib{p}')+i\frac{\epsilon_{ij\ell}}{m}(\partial_{p^{'\ell}}\Re G^R_0(\varepsilon,\mib{p}'))\right]}
  \nonumber\\
  &=&\int\frac{d\mib{p}'}{(2\pi\hbar)^d}|u_{\mib{p}-\mib{p}'}|^2\frac{\Im G^R_0(\varepsilon,\mib{p}')}{\Im\Sigma^R_0(\varepsilon,\mib{p}')}
  \nonumber\\
  &&\times\left[\frac{\hbar}{\tau_{\text{tr}}}\Sigma^<_{E_i,B_j,I}(\varepsilon,\mib{p}')+i\frac{\epsilon_{ij\ell}}{m}(\partial_{p^{'\ell}}(H_0(\mib{p}')+\Re\Sigma^R_0(\varepsilon,\mib{p}')))\right].
  \nonumber\\
  \label{eq:app:int-G^<:E,B:I:1}
\end{eqnarray}
Using Eqs.~(\ref{eq:Sigma^R:0:1}) and (\ref{eq:Sigma^<:E,B:I:1}), we reproduce Eq.~(\ref{eq:Sigma^<:E,B:I:sol}).


\begin{thebibliography}{99}

\bibitem{Mott67}
  N. F. Mott, Adv. Phys. {\bf 16} (1967), 49.

\bibitem{Ziman67}
  J. M. Ziman,  Adv. Phys. {\bf 16} (1967), 551.

\bibitem{Kubo57}
  R.~Kubo,  J. Phys. Soc. Jpn. \textbf{12} (1957), 570.

\bibitem{Nakajima58}
  S.~Nakajima,  Prog. Theor. Phys. \textbf{20} (1958), 948.

\bibitem{Zwanzig61}
  R.~Zwanzig, {\it Lectures in Theoretical Physics}, Vol. 3 (Interscience, New York, 1961).

\bibitem{Mori65}
  H.~Mori,  Prog. Theor. Phys. \textbf{33} (1965), 423;  Prog. Theor. Phys. \textbf{34} (1965), 399.

\bibitem{Fukuyama69_1}
  H.~Fukuyama, H.~Ebisawa and Y.~Wada,  Prog. Theor. Phys. \textbf{42} (1969), 494.

\bibitem{Fukuyama69_2}
  H.~Fukuyama,  Prog. Theor. Phys. \textbf{42} (1969), 1284.

\bibitem{Hurd}
  C.~M.~Hurd, {\it The Hall Effect in Metals and Alloys} (Plenum Press, New York, 1972).

\bibitem{KohnoYamada88}
  H.~Kohno and K.~Yamada,  Prog. Theor. Phys. \textbf{80} (1988), 623.

\bibitem{Keldysh}
  L.~V.~Keldysh,  Sov.\ Phys. -JETP \textbf{20} (1965), 1018 [Zh. Eksp. Teor. Fiz. \textbf{47} (1964), 1515].

\bibitem{KadanoffBaym}
  G.~Baym and L.~P.~Kadanoff,  Phys.~Rev. \textbf{124} (1961), 287.\\
  G.~Baym,  Phys.~Rev. \textbf{127} (1962), 1391.\\
  L.~P.~Kadanoff and G.~Baym, {\it Quantum Statistical Mechanics} (Benjamin, Menlo Park, 1962).

\bibitem{Altshuler78}
  B.~L.~Altshuler,  Sov.\ Phys.\ -JETP \textbf{48} (1978), 670 [Zh. Eksp. Teor. Fiz. \textbf{47} (1964), 1515].

\bibitem{RammerSmith86}
  J.~Rammer and H.~Smith,  Rev.\ Mod.\ Phys. \textbf{58} (1986), 323.

\bibitem{Mahan}
  G.~D.~Mahan, {\it Many-Particle Physics} (Plenum Press, New York, 1990), p.~671. 

\bibitem{Moyal49}
  J.~E.~Moyal,  Proc.\ Cambridge\ Philos.\ Soc. \textbf{45} (1949), 99.

\bibitem{Buttiker94}
  M.~B\"{u}ttiker, H.~Thomas and A. Pr\^{e}tre,  Z.\ Phys.\ B \textbf{94} (1994), 133.

\bibitem{LandauerButtiker}
  S.~Datta, {\it Electronic Transport in Mesoscopic Systems} (Cambridge, 1995).

\bibitem{Frensley90}
  W.~R.~Frensley,  Rev.\ Mod.\ Phys. \textbf{62} (1990), 745.

\bibitem{HaugJauho}
  H.~Haug and A.-P.~Jauho, {\it Quantum Kinetics in Transport and Optics of Semiconductors} (Springer, Berlin, 1996).

\bibitem{LiuLei}
  S.~Y.~Liu, X.~L.~Lei and N.~J.~M.~Horing,  Phys. Rev. B \textbf{73} (2006), 035323.

\bibitem{Sugimoto}
  N.~Sugimoto, S.~Onoda, S.~Murakami and N.~Nagaosa,  Phys. Rev. B \textbf{73} (2006), 113305.

\bibitem{Onoda_ahe}
  S.~Onoda, N.~Sugimoto and N.~Nagaosa, cond-mat/0605580.

\bibitem{SmrckaStreda77}
  L.~Smr\u{c}ka and P.~St\u{r}eda,  J. of Phys. C \textbf{10} (1977), 2153.

\bibitem{Streda82}
  P.~St\u{r}eda,  J. of Phys. C \textbf{15} (1982), L717

\bibitem{AGD}
  A.~A.~Abrikosov, L.~P.~Gorkov and I.~E.~Dzyaloshinski, {\it Methods of Quantum Field Theory in Statistical Physics} (Dover, New York, 1963), revised English edition translated and edited by R.~A.~Silverman.

\bibitem{SeibergWitten99}
  N.~Seiberg and E.~Witten,  J. High Energy Phys. \textbf{09} (1999), 032.

\bibitem{Non-commutativeQM}
  A.~Connes, {\it Noncommutative Geometry} (Academic, San Diego, 1994).

\bibitem{Zener}
  C.~Zener,  Proc.\ R.\ Soc.\ London A \textbf{137} (1932), 696.

\bibitem{Schwinger51}
  J.~Schwinger,  Phys. Rev. \textbf{82} (1951), 664.

\bibitem{LevandaFleurov94}
  M.~Levanda and V. Fleurov,  J. of Phys.: Cond. Mat. \textbf{6} (1994), 7889.

\bibitem{Kita01}
  T.~Kita,  Phys. Rev. B \textbf{64} (2001), 054503.

\end{thebibliography}
\end{document}